\documentclass{WileyMSP-template}

\usepackage[dvipsnames]{xcolor}
\usepackage{amsmath}
\usepackage{amsfonts}
\usepackage{graphicx}
\usepackage{amssymb}
\usepackage{siunitx} 
\usepackage{multirow}
\usepackage{todonotes}

\usepackage[normalem]{ulem}

\newcommand{\lvec}{\mathbf{l}}

\begin{document}

\pagestyle{fancy}
\rhead{\includegraphics[width=2.5cm]{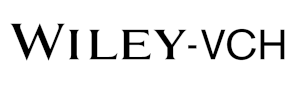}}

\title{Quantum Monte Carlo Simulation of Bipolaron Superconductivity in Extended Hubbard--Holstein models on Face-Centered-Cubic and Body-Centered-Cubic Lattices}
\maketitle

\author{G D Adebanjo}
\author{J P Hague}
\author{P E Kornilovitch}

\dedication{}

\begin{affiliations}
G D Adebanjo and J P Hague \\
School of Physical Sciences, The Open University, Walton Hall, Milton Keynes, MK7 6AA, UK \\
Email Address: Jim.Hague@open.ac.uk

P E Kornilovitch \\
Department of Physics, Oregon State University, Corvallis, OR, 97331, USA

\end{affiliations}

\date{\today}

\begin{abstract}
We investigate superlight pairing of bipolarons driven by electron-phonon interactions (EPIs) in face-center-cubic (FCC) and body-center-cubic (BCC) lattices using a continuous-time path-integral quantum Monte Carlo (QMC) algorithm. The EPIs are of the Holstein and extended Holstein types, and a Hubbard interaction is also included. Effects of adiabaticity are calculated. The number of phonons associated with the bipolaron, inverse mass, and radius are calculated and used to construct a phase diagram for bipolaron pairing (identifying the regions of pairing into intersite bipolarons and onsite bipolarons). From the inverse mass we determine that for the extended interaction, there is a region of light pairing associated with intersite bipolarons formed in both BCC and FCC lattices. Intersite bipolarons in the extended model at intermediate phonon frequency and large Coulomb repulsion become superlight due to first order hopping effects. We estimate the transition temperature, determining that intersite bipolarons are associated with regions of high transition temperatures.
\end{abstract}


\section{Introduction}

The discovery of cuprate superconductors stimulated searches for other materials that could superconduct at high temperatures. Among them are the A$_{3}$C$_{60}$ materials \cite{kratschmer1990solid}, which are doped from C$_{60}$ insulators \cite{forro2001electronic} (where A is Cs, Rb or K) and are unconventional superconductors with high transition temperatures (up to 38 K \cite{ganin2008bulk}), large Coulomb repulsion, and Jahn--Teller coupling (which has similarities to other electron-phonon couplings). The cesium-doped solid (Cs$_{3}$C$_{60}$) can have both BCC (cubic A15) and FCC structures which superconduct at similar temperatures under pressure \cite{ganin2008bulk,ganin2010polymorphism,ihara2010}. The materials have an \emph{s}-wave order parameter with signatures of unconventional superconductors such as the proximity of Mott and superconducting states \cite{alloul2013} and BCS-like theories may be insufficient to explain the superconducting mechanism in these compounds \cite{nomura2016,capone2009,gunnarsson1997_sup_in_fullerides}. As such, we are motivated to determine the properties of BCC and FCC bipolarons in systems with strong interactions between phonon modes and electrons in the presence of strong Coulomb repulsion.

One approach would be to study the simplified Hubbard-Holstein model (HHM) in which both the electron-phonon and Coulomb interactions are confined to a single lattice site \cite{Grzybowski2003,macridin2004,Grzybowski2006,Werner2007,DiFilippis2012,Kurdestany2017,Mendl2017}. HHM bipolarons form at strong electron-phonon couplings and generate lattice deformations with delta-function profiles. Coherent movement of such deformations through the lattice is exponentially suppressed. HHM bipolarons are heavy and cannot produce strong superconductivity. In the more complex but more realistic {\em extended} Hubbard-Holstein models (EHHM), the electron-phonon interaction extends beyond one site and the lattice deformation has finite dimensions \cite{Alexandrov1999,boncaehhmbipolaron,hague2006effects}. While collective energy gain from the entire deformation still must be large enough to overcome the Coulomb repulsion, the deformation on individual sites is less than in HHM, which is easier to realize. Moreover, extended lattice deformations propagate more coherently than delta-function deformations, leading to exponentially lighter EHHM bipolarons relative to the HHM case.      

In this work we will make quantum Monte Carlo simulations of the EHHM with the following form:
\begin{equation} \label{eqn:Holstein}
     H = -t\sum_{\langle ii'\rangle \sigma} c^{\dagger}_{i\sigma} c_{i'\sigma} + \hbar\omega \sum_{ij\sigma} 
     g_{ij} n_{i\sigma} \left( a_{j}^{\dagger} + a_{j} \right) + 
     \hbar\omega \sum_{j} ( a_{j}^{\dagger} a_{j} + \frac{1}{2} ) + 
     U \sum_{i} n_{i\downarrow} n_{i\uparrow},
\end{equation}
where $g_{ij}$ is a dimensionless electron phonon interaction strength, $c_{i}$ ($c^{\dagger}_{i}$) annihilate (create) electrons and the $a_{j}$ ($a^{\dagger}_{j}$) operators do the same for phonons on site $j$. $t$ is the hopping between neighboring sites. $U$ is the on site Coulomb repulsion (Hubbard $U$). The phonon frequency is $\omega$. The Holstein model corresponds to $g_{ij} = g\delta_{ij}$. The extended Holstein interaction considered here (following Ref. \cite{boncaehhmbipolaron}) only shares phonons between sites on nearest-neighbour bonds.

It can be illustrative to consider the limit of large phonon frequency, where a Lang--Firsov approximation leads to decoupling of phonons and electrons (since the average number of phonons in the wavefunction is zero), to obtain,
\begin{equation}
        \tilde{H}_{\rm LF} = -t\sum_{\langle i i'\rangle}\exp\left[-\frac{W\lambda\gamma}{\hbar\omega}\right] c^{\dagger}_{i}c_{i'} + \sum_{ii'}n_{i}n_{i'}\left(\frac{U}{2}\delta_{ii'}-\frac{W\lambda\Phi_{ii'}}{\Phi_{00}}\right) - \sum_{i}Un_{i}+ \hbar\omega\sum_{j}\left(a^{\dagger}_{j}a_{j}+\frac{1}{2}\right)
\end{equation}
where the effective phonon-mediated interaction has the form,
\begin{equation}
\Phi_{ii'} = \hbar\omega \sum_{j}g_{ij}g_{ji'},
\end{equation}
and we used $\Phi_{\rm NN}$ to denote near-neighbour phonon-mediated coupling. Here, the non-interacting half bandwidth, $W=zt$, where $z$ is the number of near neighbours and we defined $\gamma=1-\Psi_{\rm NN}/\Phi_{00}$. For the Holstein and extended Holstein interactions considered, $\Phi_{ii'}$ is only non-zero for on-site and near-neighbor sites. The electron-phonon interaction strength is,
\begin{equation}
    \lambda = \frac{\Phi_{00}}{W} \: .
\end{equation}
Thus the effective on-site and inter-site coupling strengths at large phonon frequency are $U' = U-2W\lambda$ and $V' = -2W\lambda\Phi_{\rm NN}/\Phi_{00}$.

\begin{figure}
\centering
\includegraphics[width=0.7\textwidth]{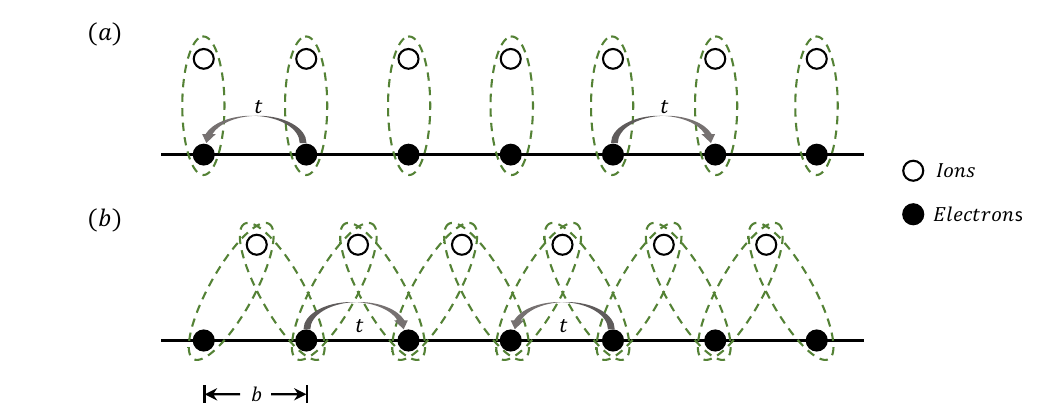}
\renewcommand\baselinestretch{1}\selectfont
\caption{One-dimensional schematic of (a) the Holstein-Hubbard model and (b) the extended Holstein-Hubbard model with near-neighbour EPI introduced in Ref. \cite{boncaehhmbipolaron} and studied here. The filled circles, empty circles and dashed oval circles represent the electron Wannier orbitals, lattice ions, and nonzero electron-phonon coupling, respectively. The nearest-neighbour electron sites have overlapping orbitals such that an electron can hop via $t$, and $b$ is the intersite spacing.}
\label{fig:hubbard_holstein_and_bonca_EPI_schematic}
\end{figure}

Bipolarons formed in the Holstein-Hubbard model (HHM) and those formed in a truncated EHHM \cite{boncaehhmbipolaron} consist of two polarons that can bind into a stable pair when the exchange of phonons overcomes Coulomb repulsion. These categories of bipolarons differ in the spatial extent of their EPI. Models of this $U-V$ type have been studied in the anti-adiabatic limit \cite{micnas1990,Kornilovitch2024}. As shown schematically in Figure \ref{fig:hubbard_holstein_and_bonca_EPI_schematic}, the nature of the EPI in the HHM is site-local (electron located at the vibrating atomic site) whereas the EHHM has a long-range EPI (of near-neighbour type). To our knowledge, there are no numerical studies yet of bipolaron properties in either the BCC or the FCC lattice in the adiabatic limit of low phonon frequency. We note our previous QMC studies of bipolarons on the chain \cite{hague2009bipolarons_sing_trip_chain}, square lattice \cite{hague2010light}, triangular lattice \cite{hague_staggered_ladder,hague2010light} and simple cubic lattice \cite{davenport2012}.

The transition temperature of a superconductor of real-space bipolarons in the BEC regime depends critically on their mass and size. In the dilute limit, where there is infrequent scattering {\em between} bipolarons, the transition temperature has the form,
\begin{align}
	T_{\rm BEC} & = \frac{3.31 \hbar^{2} n^{2/3}_{b} }{k_{B} m^{\ast\ast}} \: ,
 \label{eqn:bectc}
\end{align}
where $m^{\ast\ast}$ is the bipolaron (pair) mass, and $n_{b}$ is their density \cite{alexandrov1994}. As $n_{b}$ increases, $T_{\rm BEC}$ also increases until bipolarons start to overlap, at which point the transition temperature peaks before decreasing. Saturation of $T_{\rm BEC}$ is estimated to occur in the vicinity of {\em close packing}. Thus, a close packing transition temperature is defined to be \cite{Ivanov1994,Kornilovitch2015,Kornilovitch2024,adebanjo2024}
\begin{equation}
\frac{k_{B} T^{\ast}}{t} \sim \frac{6.62}{ \Omega_{p}^{2/3} } \frac{m_0}{m^{\ast\ast}} ,
 \label{eqn:twoprime}
\end{equation}
where the bare {\em electron} mass is $m_{0} = \hbar^2/(2t a^2)$, and $\Omega_{p}$ is the bipolaron volume in units of the lattice constant $a$. $\Omega_{p}$ needs to be small enough such that inter-bipolaron scattering does not significantly affect Eq.~(\ref{eqn:bectc}). We calculate $\Omega_{p} = 4\pi R^{\prime 3} / 3$, where $R' = R$ (if $R>b$) and $R=b$ otherwise (since small bipolarons have hard cores).

\begin{figure*}[t]
    \centering
    \includegraphics[width=140mm]{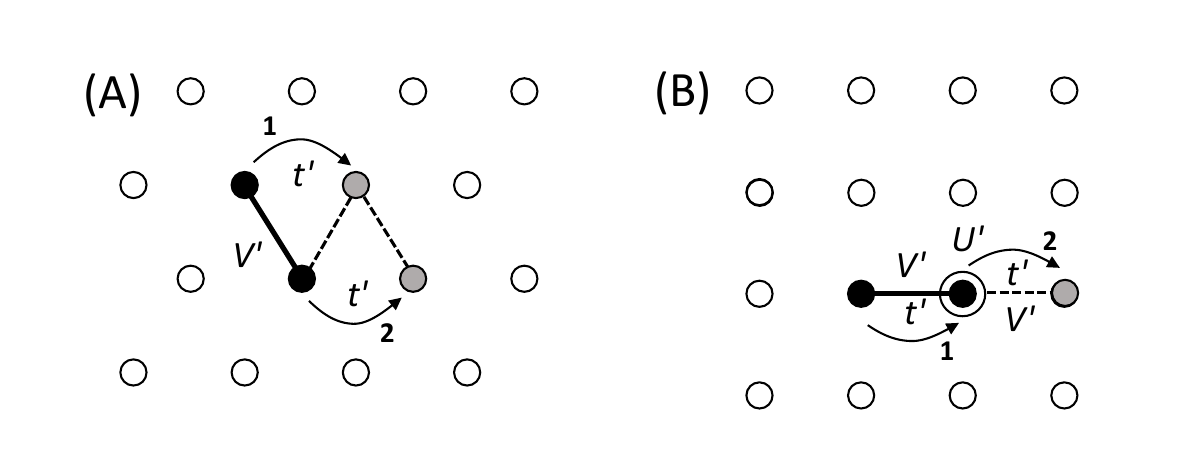}
    \caption{Schematics of different light pair mechanisms. (A) On the triangular lattice, a pair with NN attraction $V'$ moves in the first order of NN hopping $t$. (B) In the resonant case, $V' = U'$, the pair can move in the first order of NN hopping $t$ even on the square lattice. Numbers `1' and `2' indicate hopping order.}
    \label{fig:schematic}
\end{figure*}

We now discuss applicability of the ideal-gas BEC formula, Eqs.~(\ref{eqn:bectc},\ref{eqn:twoprime}), to fullerene physics and to superconductivity in general. Clearly, Eqs.~(\ref{eqn:bectc},\ref{eqn:twoprime}) neglect the interaction between bipolarons and as such provide only an estimate for the transition temperature. At the same time, in the well-documented case of superfluid helium-4, which is arguably a strongly interacting system, the $T_{c}$ error introduced by neglecting the interaction is only about 40\%. We therefore believe that Eqs.~(\ref{eqn:bectc},\ref{eqn:twoprime}) are adequate to investigate qualitative trends and relative magnitudes of superconducting effects, which is the focus of this paper. Ideally we would solve the full many-body problem, Eq.~(\ref{eqn:Holstein}), at finite electron densities and without approximations, but it is still not feasible. Path-Integral QMC employed here is among the most powerful unbiased methods to study many-body Hamiltonians, but its application to the many-fermion sector of Eq.~(\ref{eqn:Holstein}) is severely limited by the sign problem. We also note that in strongly interacting electron-phonon systems, even the two-fermion problem, i.e., the bipolaron, is highly complex and nontrivial. Our approach to superconductivity in this and related papers has been to investigate the two-fermion problem without approximations and then to accept the $< 40\%$ absolute error in $T_c$ which still enables a meaningful analysis of relative effects. In practice, we expect this error to be much lower, since estimates of corrections to $T_{c}$ imply a fractional reduction proportional to $n^{2/3}$ for hard-core bosons and proportional to $n$ for inter-boson interactions  \cite{micnas1990,alexandrov1986}. The remarkable feature of Eq.~(\ref{eqn:twoprime}) is that it contains only two-fermion properties and immediately provides physical insights whenever they are known.

As can be appreciated from Eqs.~(\ref{eqn:bectc}) and (\ref{eqn:twoprime}), the critical temperature is inversely proportional to the bipolaron mass. Thus in the context of superconductivity, it is important to understand when bipolarons are heavy and when they may be light. The range of electron-phonon interaction affects $m^{\ast\ast}$ in two crucial ways. First, it affects the {\em polaron} mass (i.e. the mass of the two bipolaron constituents), $m^{\ast} \sim  m_0 \exp{(\gamma W \lambda/\hbar\omega)}$ (in the limit of large $\hbar\omega/W$). For the local interaction of the HHM, the lattice deformation is confined to the site occupied by the fermion and is very localised. When the polaron moves, the entire deformation must move to the neighboring site resulting in an exponentially large $m^{\ast} \sim m_0 \exp{( W \lambda/\hbar\omega)}$, since $\gamma=1$ for the HHM. In the EHHM, the lattice on the nearest site is partially pre-deformed. {\em Additional}  deformation associated with polaron movement is smaller than in the Holstein case, which leads to $\gamma < 1$. Thus, EHHM polarons can be {\em exponentially} lighter than HHM polarons. We call such polarons ``light'' \cite{hague2007superlight}.

The second effect of the interaction range is related to bipolaron movement. In the HHM, both polarons occupy the same lattice site in the S0 (on-site singlet) configuration. Bipolaron movement is a second-order process that involves an intermediate state with energy excess equal to the bipolaron binding energy $\Delta$. As a result, the bipolaron mass scales as $m^{\ast\ast} \propto (m^{\ast})^2 \Delta$ in the anti-adiabatic limit. Holstein bipolarons are quadratically heavier than Holstein polarons. This result applies to Holstein models in lattices of all dimensionalities. The same scaling $m^{\ast\ast} \propto (m^{\ast})^2 \Delta$ applies to extended EHHM bipolarons on the BCC lattice, although in this case the constituent polarons are light and therefore the bipolaron is also light. However, EHHM bipolarons on the FCC lattice may display a qualitatively different behavior. If the Hubbard repulsion is strong enough, the two polarons will be forced to occupy two neighbouring sites in S1 configuration. Due to the geometry of the FCC lattice, the S1 bipolaron can move in the {\em first order} in polaron hopping in the large $\omega$ limit, as illustrated in Figure \ref{fig:schematic}. The S1 bipolaron mass scales as $m^{\ast\ast} \propto m^{\ast}$. We call such bipolarons ``superlight'' \cite{hague2007superlight}. 

We expect, therefore, all HHM bipolarons to be very heavy and suppress $T^{\ast}$. The EHHM bipolarons on the BCC lattice are expected to be light with an intermediate $T^{\ast}$. On the FCC lattice, the EHHM bipolarons are light in the S0 configuration and superlight in the S1 (intersite singlet) configuration. The latter provides the highest $T^{\ast}$.

The goal of this study is to understand whether the above qualitative picture is maintained as $\omega$ is lowered towards the adiabatic limit using a numerically exact QMC simulation method. The paper is organised as follows: In Section \ref{sec:method} an overview of the quantum Monte Carlo scheme is given, in addition to introducing a new triple update needed when simulating FCC lattices in Section \ref{sec:triple_update}. In Section \ref{sec:results} results from the simulations are presented. Finally, in Section \ref{sec:conclusions} a summary can be found.

\section{Method}
\label{sec:method}

In this work, we use a continuous-time path-integral quantum Monte Carlo scheme to simulate the HHM and EHHM defined in Eq.~(\ref{eqn:Holstein}) (see e.g. Refs. \cite{hague2007superlight,hague2010light,davenport2012}). We place two fermions in a $20 \times20 \times 20$ lattice constant box, and then carry out simulations for $\bar{\beta} = t/k_{B}T = 20$ (where $k_{B}$ is Boltzmann's constant and $T$ the temperature) at phonon frequencies corresponding to $\hbar\omega/t = 1$ and $\hbar \omega/W = 1$. The Hamiltonian is simulated for a range of $U$ and $\lambda$. We adopt twisted boundary conditions on path ends to estimate the mass. Detailed procedures for the simulation and evaluation of estimators for a range of bipolaron properties have been discussed in earlier papers (see Refs. \cite{hague2009bipolarons_sing_trip_chain,hague2010light,davenport2012}, and references therein). Measurements are made every few Monte Carlo steps.

For BCC lattices, we use the update scheme summarised in Sec. \ref{sec:mainmethod} (which is essentially the same set of updates as for simple cubic lattices detailed in Ref. \cite{davenport2012}). For FCC lattices, these updates plus an additional update analogous to the triple update used to simulate bipolarons on triangular lattices (Sec. \ref{sec:triple_update}) are needed (see Ref. \cite{hague2010light}).

\subsection{QMC Update Rules and Weighting Scheme}
\label{sec:mainmethod}

The QMC method used is a continuous-time path integral approach. The electron paths are continuous in time and inter-site electron hops appear as kinks in the paths. We only studied singlet states, meaning that there is no sign problem, since both direct and exchanged configurations are sign positive. Kink insertion or removal is carried out in pairs to ensure that twisted boundary conditions are respected (i.e. the end configurations of the paths are in the same relative positions up to a translation). Following our established algorithm (see Refs. \cite{davenport2012,hague2009bipolarons_sing_trip_chain,hague2010light}), four binary updates were used: (1) A kink of type $\lvec$ is added to (or removed from) each of both paths. (2) A kink of type $\lvec$ and its anti-kink $-\lvec$ are added to (removed from) one of the paths. (3) A kink $\lvec$ and antikink $-\lvec$ are added to (or removed from) different paths (4) A pair of kinks of type $\lvec$ is added to (or removed from) one of the paths. 

On insertion or removal, kinks can be shifted at larger imaginary times (top shift in direction $\lvec$) or at earlier imaginary times (bottom shift in direction $-\lvec$). Here, $\lvec$ is a near-neighbor vector.

In all cases, a primary path (denoted $A$) and kink type are selected with probability $1/2$ and $P_{\lvec}=1/N_{k}$ where $N_{k}$ is the number of nearest neighbours. The other path is denoted $B$. Where kink insertion is chosen, one (two) imaginary time(s) $\tau$ is (are) selected from a uniform probability density $p(\tau)=1/\beta$. For kink removal, a kink of type $\lvec$ is chosen from path $X$ with probability $1/N_{X\lvec}$ where $X$ represents path $A$ or $B$.

In the binary updates, correlated shifts indicate that e.g. both kinks relate to a top shift, whereas anti-correlated shifts that each kink comes with a different shift. Correlation and anti-correlation cases are considered to be different subtypes of the update. In the direct configuration, only updates (1) and (2) are available and both correlated and anti-correlated shifts are allowed. In the exchanged configuration, updates (1) and (2) are available as correlated shifts only and (3) and (4) as anti-correlated shifts only. Thus there are four update subtypes attempted in the direct or exchanged configuration. When the path ends are both on the same site, all eight update subtypes are possible. In general, correlated updates maintain the distance between paths and anti-correlated updates change the distance between paths.

Following the update, the electron-phonon action is calculated (see Ref. \cite{hague2007superlight}),
\begin{eqnarray}
A[{\bf r}(\tau)] & =& \frac{z\lambda\bar{\omega}}{2\Phi_0(0,0)}
\int_0^{\bar\beta} \int_0^{\bar\beta} d \tau d \tau'
e^{-\bar{\omega} \bar\beta/2}  \sum_{ij}\Phi_0[\mathbf{r}_i(\tau),\mathbf{r}_j(\tau')] \left( e^{\bar{\omega}(\bar\beta/2-|\tau-\tau'|)} + e^{-\bar{\omega}(\bar\beta/2-|\tau-\tau'|)} \right)\nonumber\\
 & & + \frac{z\lambda\bar{\omega}}{\Phi_0(0,0)}
 \int_0^{\bar\beta} \int_0^{\bar\beta} d \tau d \tau' e^{- \bar{\omega} \tau}
 e^{-\bar{\omega}( \bar\beta - \tau')} \sum_{ij}\left( \Phi_{\Delta\mathbf{r}}[\mathbf{r}_i(\tau),\mathbf{r}_j(\tau')] -
 \Phi_0[\mathbf{r}_i(\tau),\mathbf{r}_j(\tau')]\right)\nonumber\\
& & - \frac{U}{2}\int_0^{\beta}\delta_{\mathbf{r}_1(\tau),\mathbf{r}_2(\tau)}\,d\tau
\label{eq:action}
\end{eqnarray}
where $\bar{\omega}=\hbar\omega/t$, $\bar{\beta}=t/k_{B}T$, $\Delta \mathbf{r}$ is the twist in the boundary conditions in time and $z$ the number of nearest neighbours.

The detailed balance equations and Metropolis-Rosenbluth conditions for acceptance and rejection of these updates can be found in Ref. \cite{hague2007superlight}. A new update rule for FCC lattices essential for ergodicity is detailed in the next subsection. If a kink removal update is selected and can't be carried out because there are insufficient kinks of the correct type to remove, then the update is rejected.

\subsection{Three-kink update for FCC Lattice}\label{sec:triple_update}

\begin{figure}
\centering
    \includegraphics[width=0.4\textwidth]{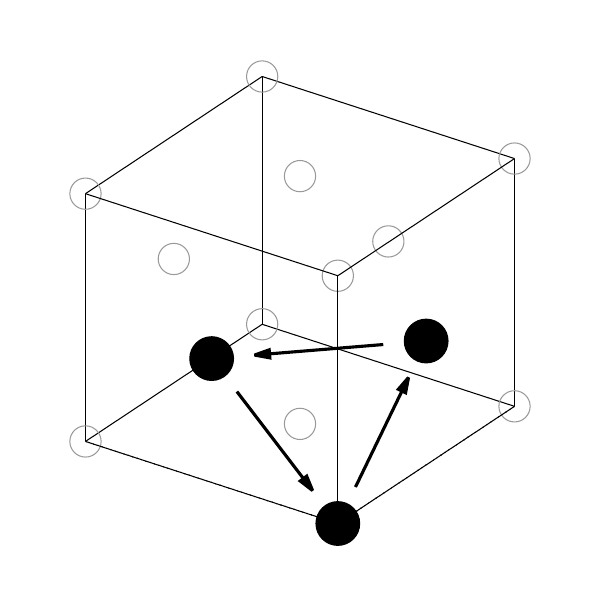}
    \caption{Schematic showing a three-hop loop on an FCC lattice.}
    \label{fig:threehop}
\end{figure}

Analogously to the case of the triangular lattice \cite{hague2010light}, electrons on an FCC lattice can return to their original position through three hops. This means that part of the configuration space for the bipolaron containing paths with odd numbers of kinks is not accessible through binary updates. This is generally true for any Hamiltonian where three kinks can be selected such that, $\lvec_{1}+\lvec_{2}+\lvec_{3}=0$. For simple lattices with nearest neighbor hops only, this includes only the triangular and FCC cases. Hence, at lease one three-kink update must be included in the Monte Carlo scheme when simulating FCC lattices to ensure ergodicity. Several three-kink updates are possible in principle, but in practice, an update introducing three kinks to a single path is sufficient to ensure ergodicity. An example of a three-hop loop is shown in Fig. \ref{fig:threehop}.

A set of three kinks is selected with the property $\lvec_{1}+\lvec_{2}+\lvec_{3}=0$ with equal probability from all possible sets. The update then attempts to insert or remove these three kinks with equal probability onto one of the paths. As before, where kink insertion is chosen, their imaginary times, $\tau_1, \tau_2$ and $\tau_3$ are selected from a uniform probability density $p(\tau)=1/\beta$. For kink removal, a kink of type $\lvec$ is chosen with probability $1/N_{A\lvec}$.

The Metropolis condition for insertions is,
\begin{equation}\label{eq:accept_prob_insertion}
        P_{\rm add} = {\rm  min} \left\{ 1, \frac{\bar{\beta}^3}{(N_{\lvec_{1}}+1)(N_{\lvec_{2}}+1)(N_{\lvec_{3}}+1)} e^{A_{f}-A_{i}}\right\}
    \end{equation}
and for removals,
\begin{equation}\label{eq:accept_prob_removal}
        P_{\rm remove} = {\rm  min} \left\{ 1, \frac{N_{\lvec_{1}}N_{\lvec_{2}}N_{\lvec_{3}}}{\bar{\beta}^3} e^{A_{f}-A_{i}} \right\}
\end{equation}
unless the path contains no kinks of type $\lvec_1, \lvec_2$ or $\lvec_3$, in which case the update is rejected. Here $N_{\lvec_1}$ etc. are the number of kinks in the initial configuration, $A_{i}$ is the initial action and $A_{f}$ the final action. Further details on triple updates can be found in Refs. \cite{ganiyuthesis} and \cite{hague2010light}.

\section{Results} \label{sec:results}

This section describes singlet bipolarons in both the HHM and EHHM. Properties of the bipolarons are computed for both BCC and FCC lattices, and include the ground state energy, the total number of excited phonons, the effective mass, and the bipolaron radius. From these, the phase diagram is determined.

\begin{figure}
\includegraphics[width=0.3\textwidth]{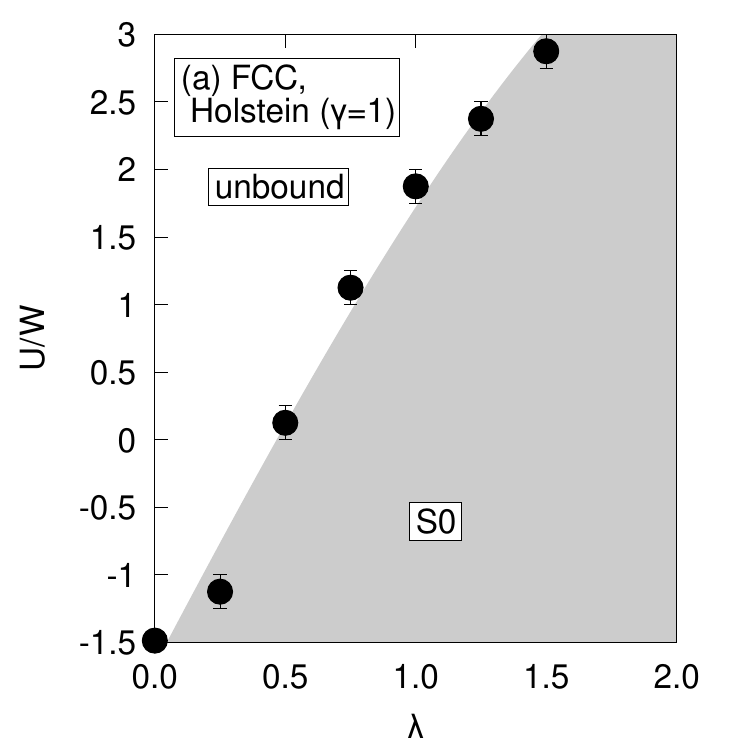}
\includegraphics[width=0.3\textwidth]{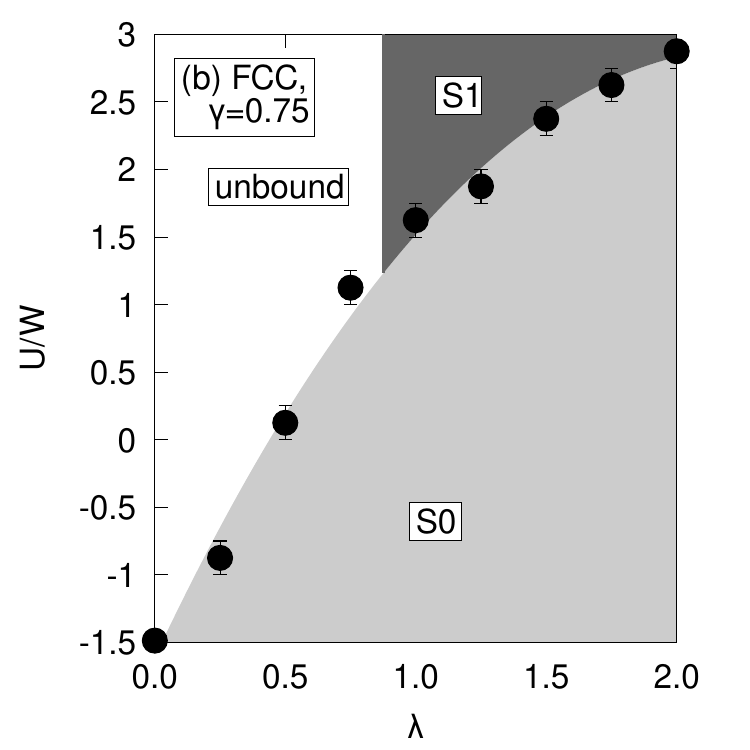}
\includegraphics[width=0.3\textwidth]{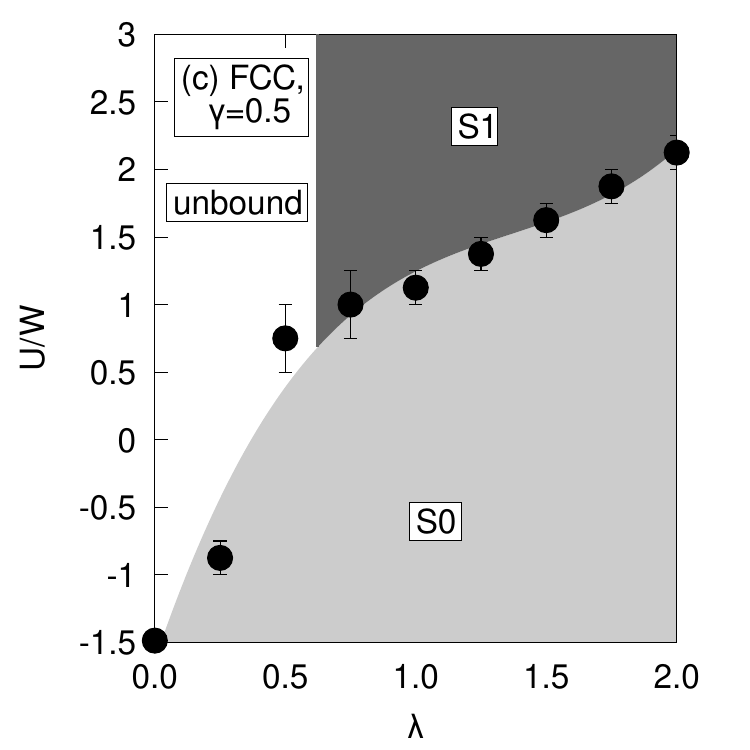}
\includegraphics[width=0.3\textwidth]{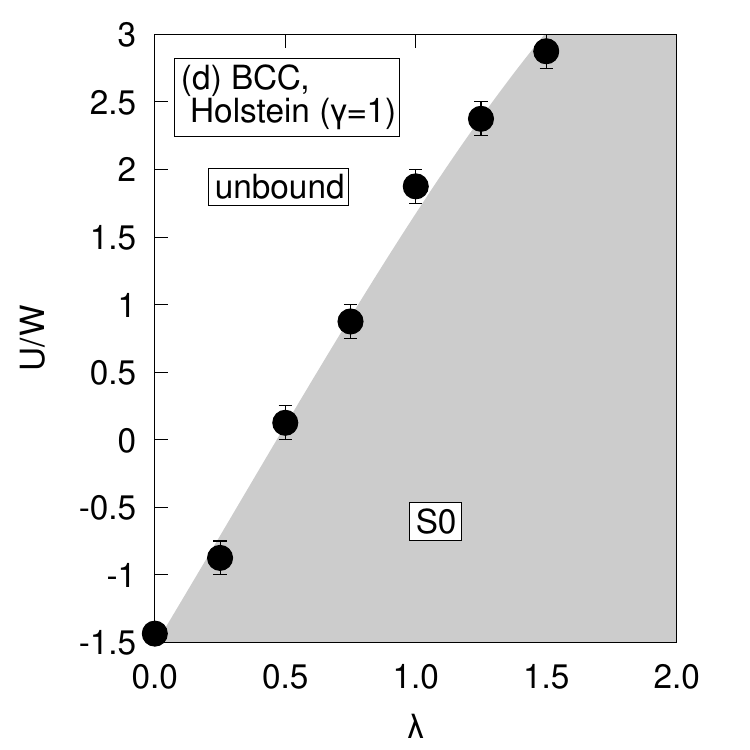}
\includegraphics[width=0.3\textwidth]{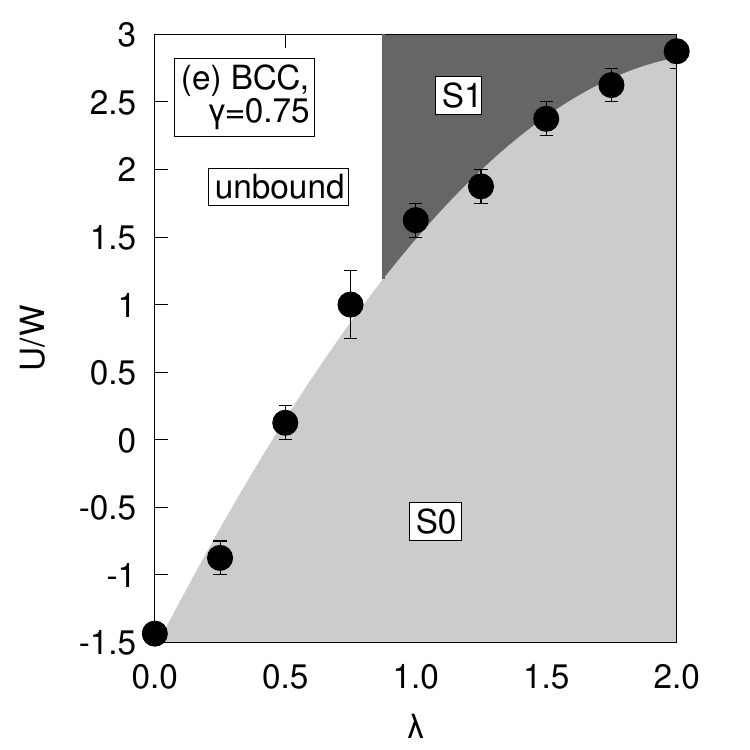}
\includegraphics[width=0.3\textwidth]{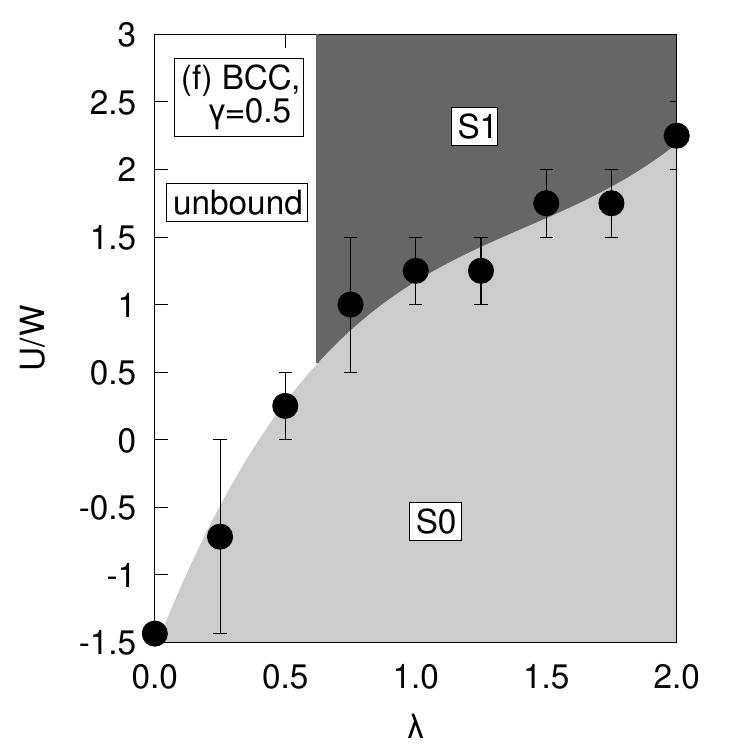}
    \caption{Phase diagram of the extended Holstein model on BCC and FCC lattices shown for three different values of $\gamma=1-\Phi_{\rm NN}/\Phi_{00}$. Calculations are in the adiabatic limit, $\hbar\omega = t$, which corresponds to $\hbar\omega = W/12$ for BCC and $W/16$ for FCC cases. The additional intersite interaction promotes S1 pairing leading to an increase in the size of the S1 region such that it can be found at lower $U$ and $\lambda$. The case $\gamma=1$ corresponds to the Holstein interaction. Lines are a guide to the eye, determined by fitting a cubic curve to the points defining the S0 boundary.}
    \label{fig:phasediagram}
\end{figure}

\subsection{Adiabatic limit}

In this section, the adiabatic limit of, $\hbar\omega = t$ is explored, which corresponds to $\hbar\omega = W/12$ for BCC and $W/16$ for FCC cases. 

The phase (binding) diagram of the Holstein and extended Holstein bipolarons is shown in Fig. \ref{fig:phasediagram}.  The S0 part of the phase diagram was determined by searching for a sudden drop in the number of phonons associated with the two particles as $U$ is increased at fixed $\lambda$ followed by a region where there is no change in $N_{\rm ph}$ on further increase in $U$. For the regions of parameter space with no S0 bipolaron, the S1 bipolaron is identified by determining where the particle separation is the intersite spacing, $b$ and unbound cases where the particle separation is very large. For FCC lattice, $b=a/\sqrt{2}$ and for BCC lattice, $b=a\sqrt{3}/2$. The Holstein cases are shown in panels (a) and (d). We have not identified an S1 region in this case, although we note that a tiny S1 region can be observed for the 2D Holstein bipolaron \cite{macridin2004}. 

\begin{figure}
\includegraphics[width=0.3\textwidth]{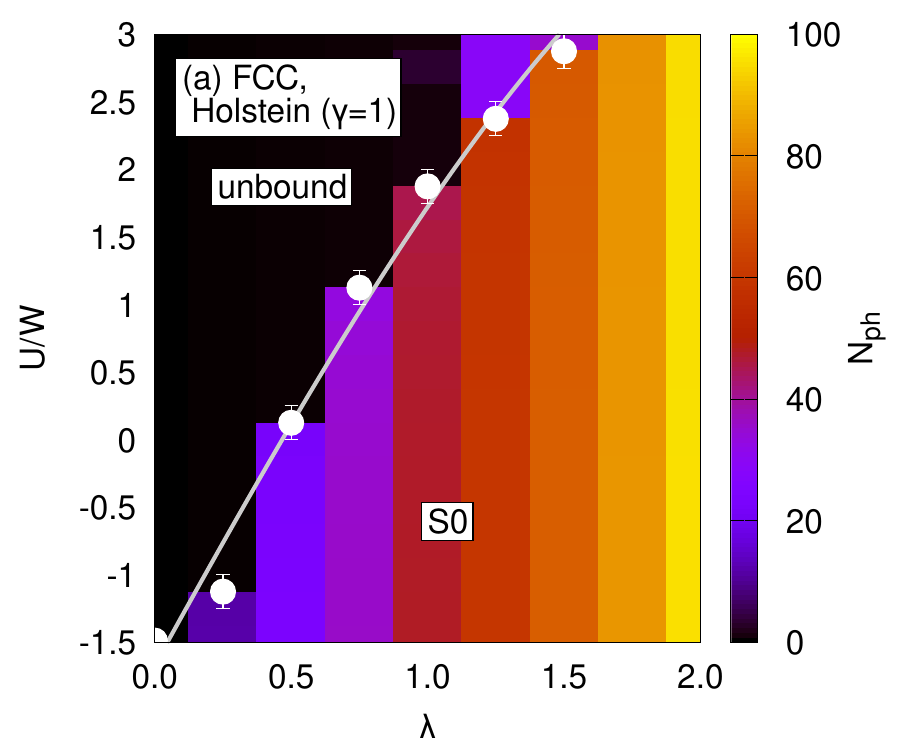}
\includegraphics[width=0.3\textwidth]{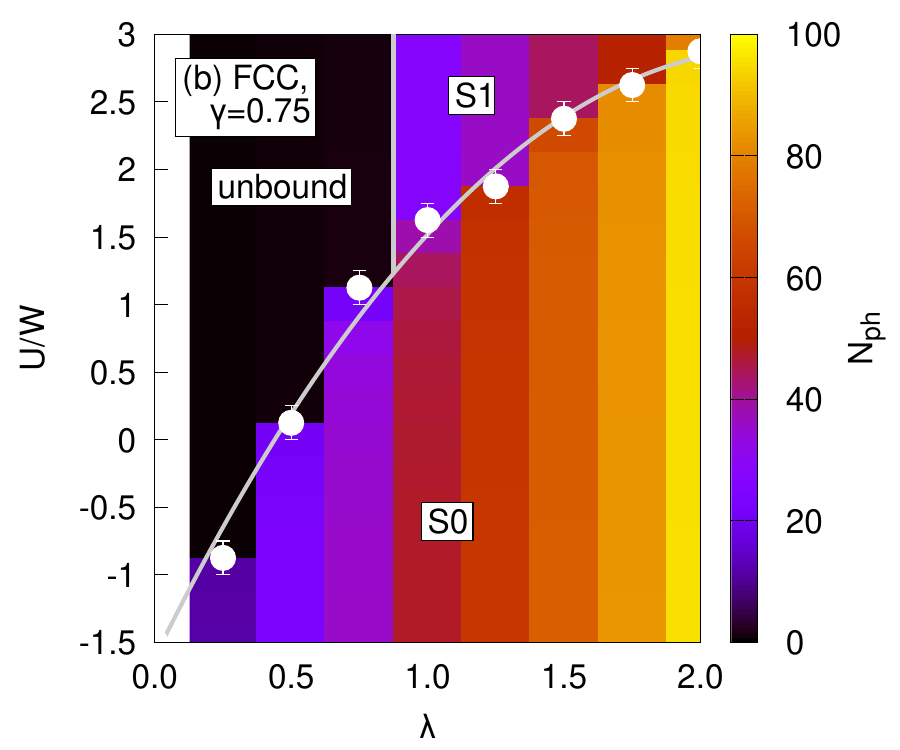}
\includegraphics[width=0.3\textwidth]{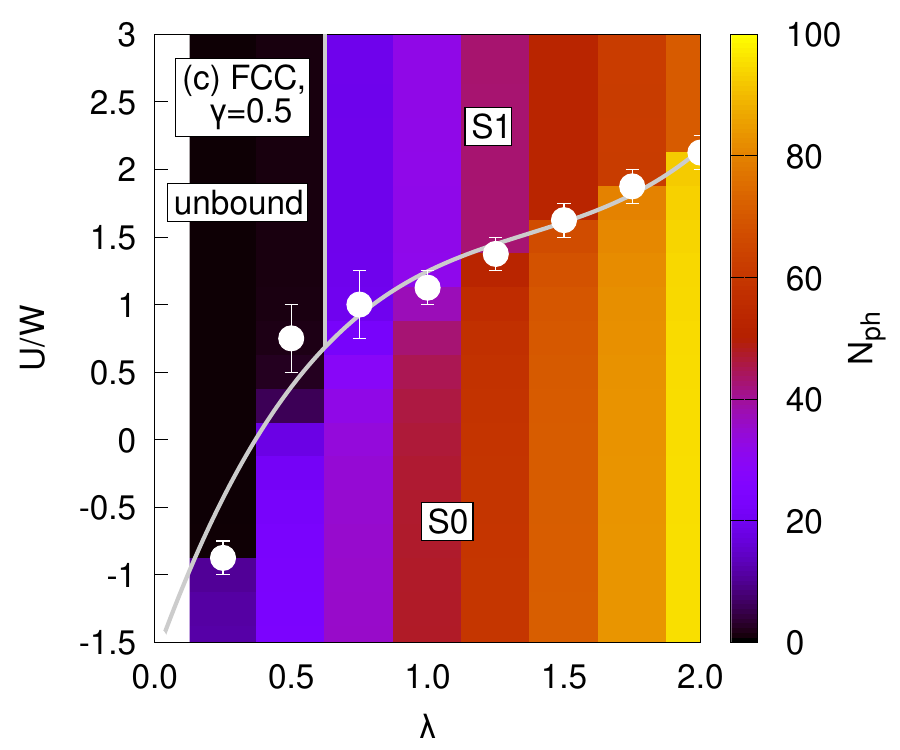}
\includegraphics[width=0.3\textwidth]{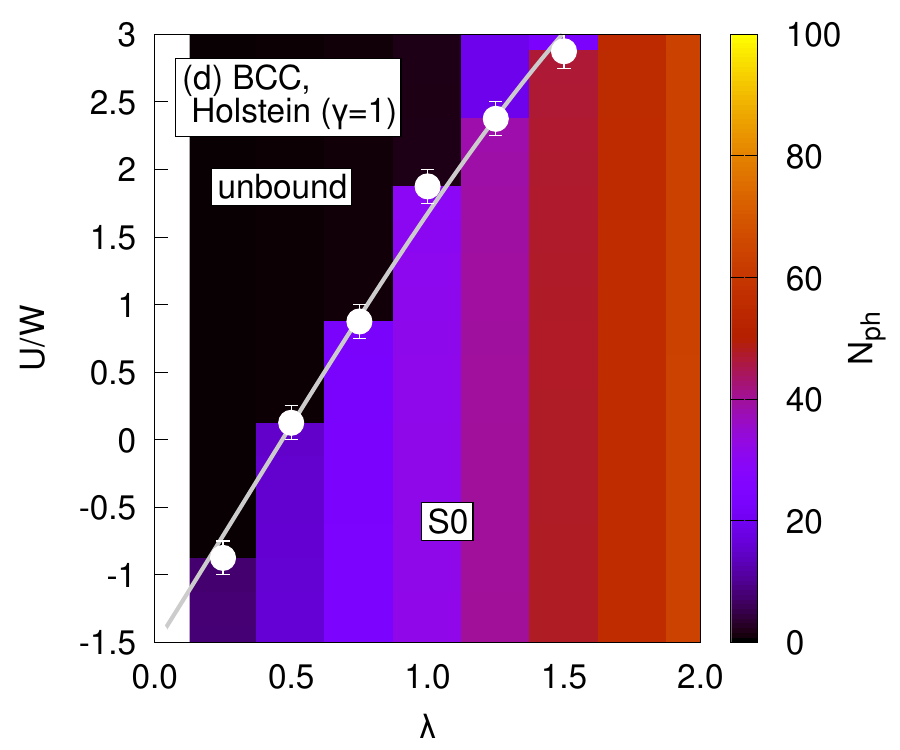}
\includegraphics[width=0.3\textwidth]{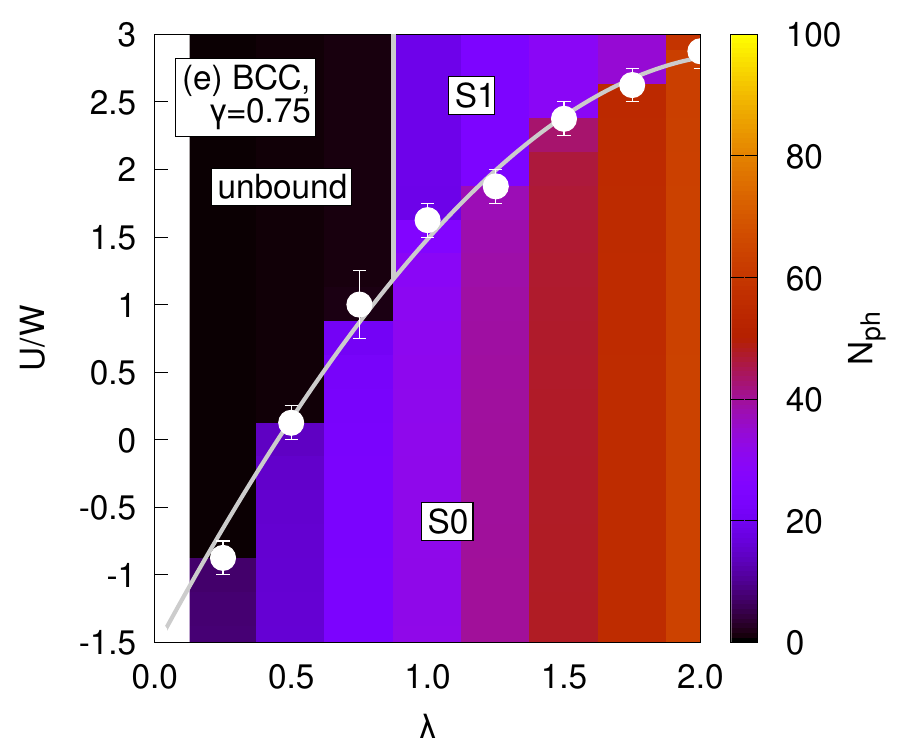}
\includegraphics[width=0.3\textwidth]{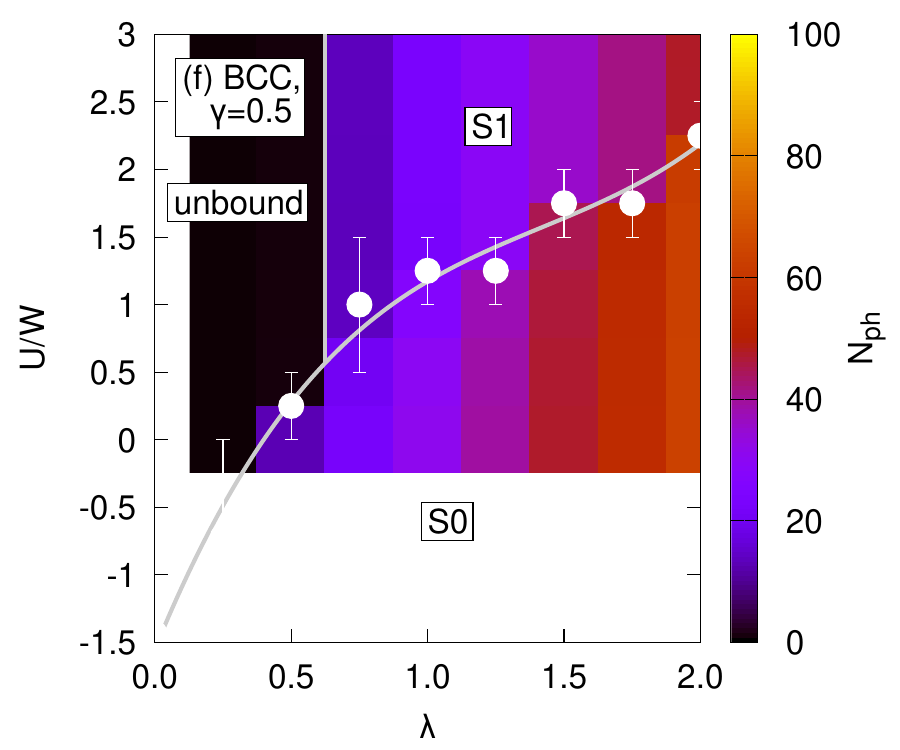}
    \caption{Number of phonons associated with the bipolaron. Calculations are in the adiabatic limit, $\hbar\omega = t$, which corresponds to $\hbar\omega = W/12$ for BCC and $W/16$ for FCC cases. An abrupt change in the number of phonons corresponds to the boundary between S0 and S1 bipolarons. There are approximately 50\% more phonons associated with the FCC lattice due to the larger kinetic energy relative to $\hbar\omega$ in that case. $N_{\rm ph}$ is constant on change of $U$ for S1 and unbound cases.}
    \label{fig:noofphonon}
\end{figure}

An abrupt change in the number of phonons, $N_{\rm ph}$, associated with the bipolaron can be found at the boundary of the region of parameter space containing S0 states (Fig. \ref{fig:noofphonon}). For unbound states and S1 states, $N_{\rm ph}$ and other properties do not vary on change of the site local Hubbard $U$. For the unbound case, the abrupt change occurs because there is a smaller phonon cloud associated with a polaron than a bipolaron (the bipolaron self reinforces the phonon cloud). There are also fewer phonons associated with the S1 bipolaron, since the electron-phonon interaction is stronger on-site than intersite, so there is less opportunity to create phonons.

\begin{figure}
\includegraphics[width=0.3\textwidth]{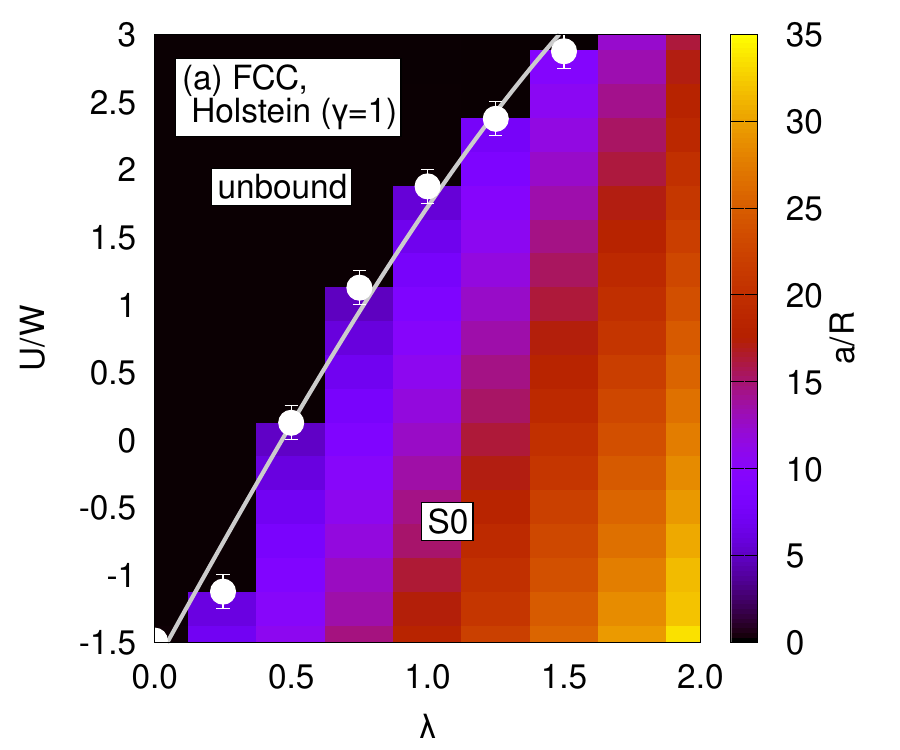}
\includegraphics[width=0.3\textwidth]{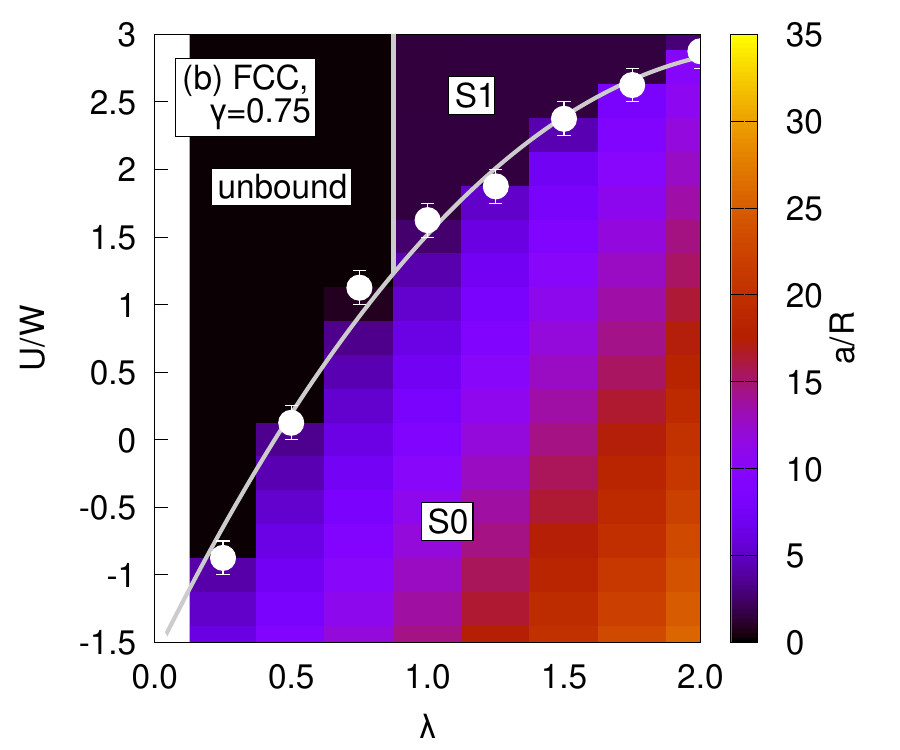}
\includegraphics[width=0.3\textwidth]{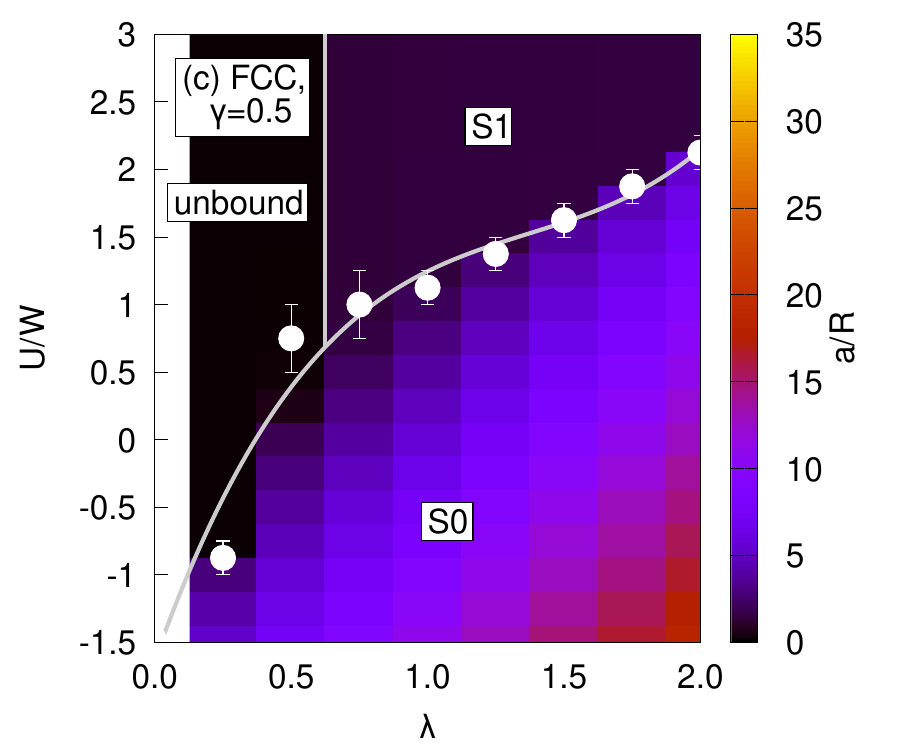}
\includegraphics[width=0.3\textwidth]{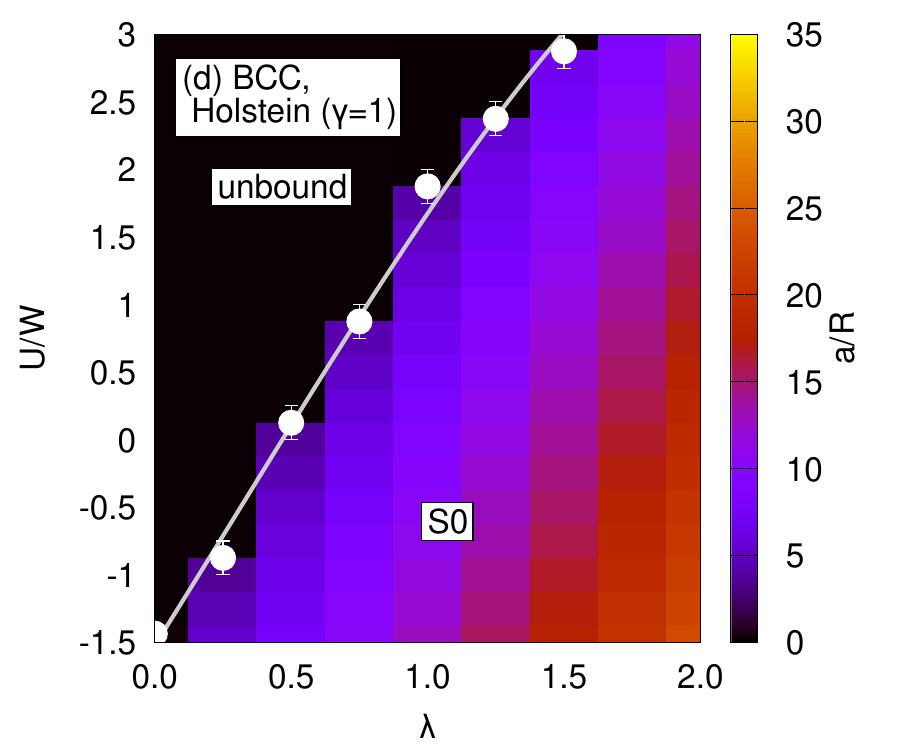}
\includegraphics[width=0.3\textwidth]{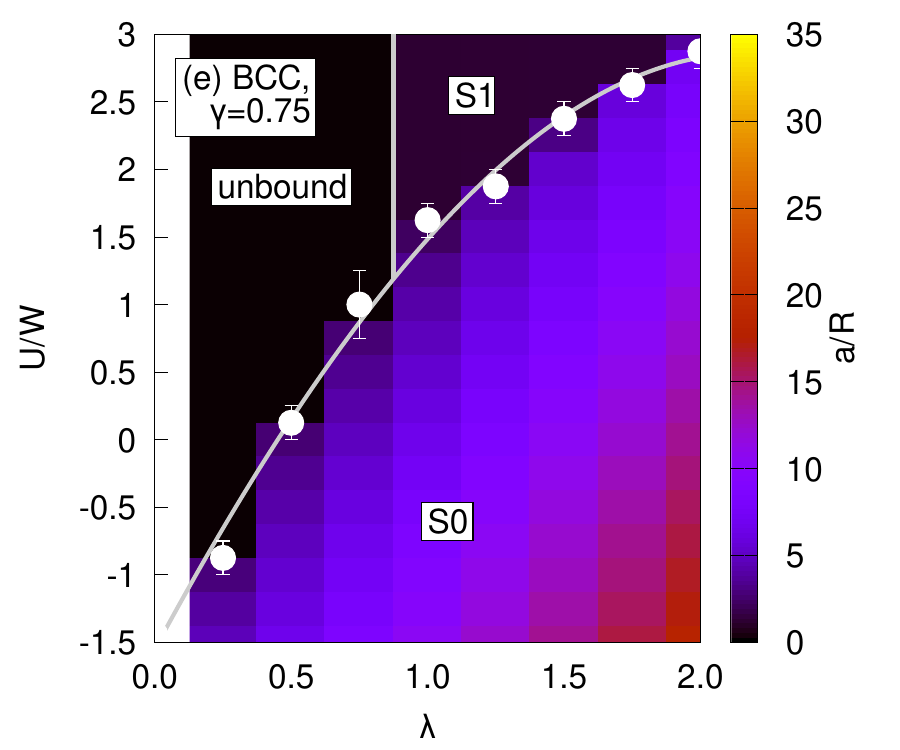}
\includegraphics[width=0.3\textwidth]{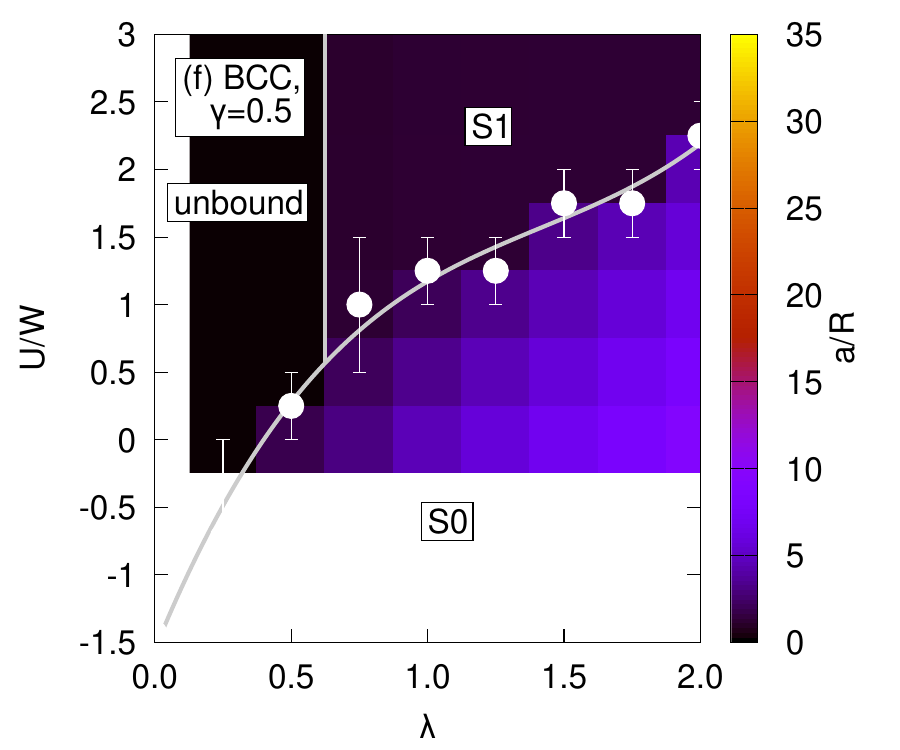}
    \caption{Inverse radius of the bipolaron, $a/R$. The S1 bipolaron is characterized by pairs with interparticle spacing, $b$ at large $U$ and large $\lambda$. Calculations are in the adiabatic limit, $\hbar\omega = t$, which corresponds to $\hbar\omega = W/12$ for BCC and $W/16$ for FCC cases. The S0 bipolaron is small and characterized by large inverse radius (found for large $\lambda$ and small $U$). Regions with no pairing are found for large $U$ and small $\lambda$ and shown in black.}
    \label{fig:invradius}
\end{figure}

Figure \ref{fig:invradius} shows the inverse radius of the bipolaron. Again computations are shown for BCC and FCC lattices and a range of $\gamma$ is considered. The regions of the parameter space where S1 bipolaron are formed can be determined by finding pairs with interparticle spacing, $b$. This region can be identified as an area of constant $a/R$ in the upper right corner of the plots for large $U$ and large $\lambda$. The S0 bipolaron is smaller than the intersite spacing showing as a large inverse radius (found for large $\lambda$ and small $U$ such that the electron-phonon coupling overcomes the Coulumb repulsion).  Regions with no pairing are found for large $U$ and small $\lambda$ such that Coulomb repulsion overcomes the electron-phonon coupling and are shown in black. We note that inverse radius is displayed so that the unbound regions (with large radius) do not dominate the plot.

\begin{figure}
\includegraphics[width=0.3\textwidth]{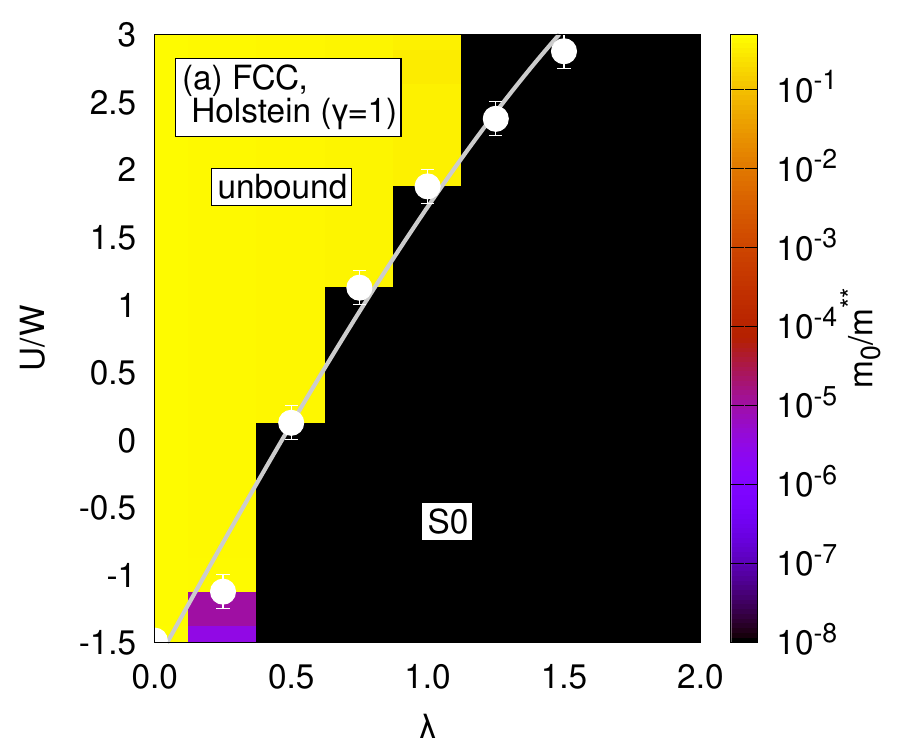}
\includegraphics[width=0.3\textwidth]{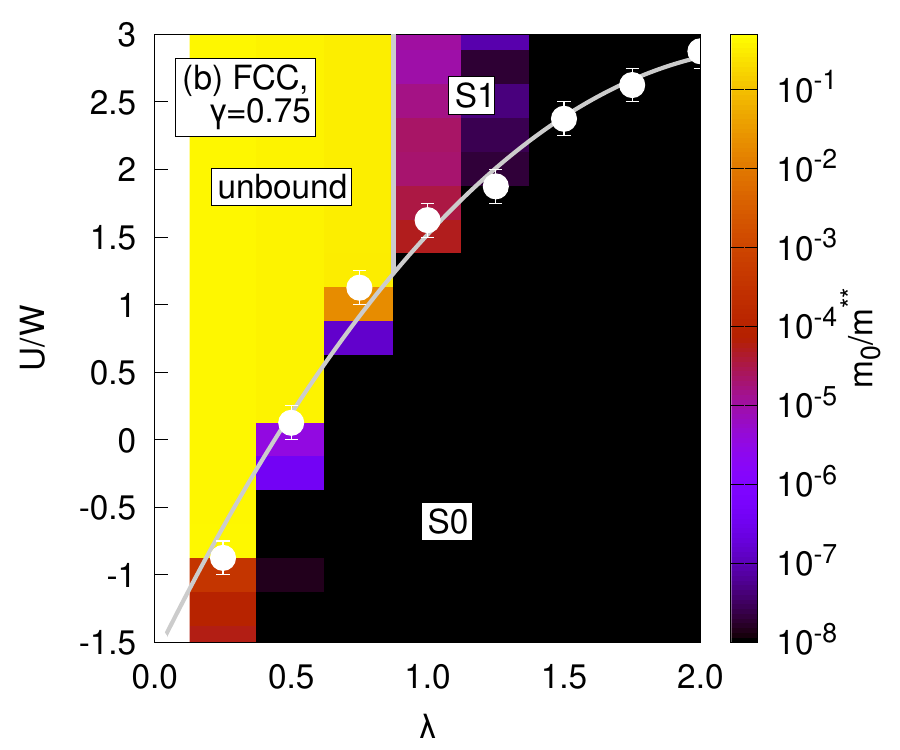}
\includegraphics[width=0.3\textwidth]{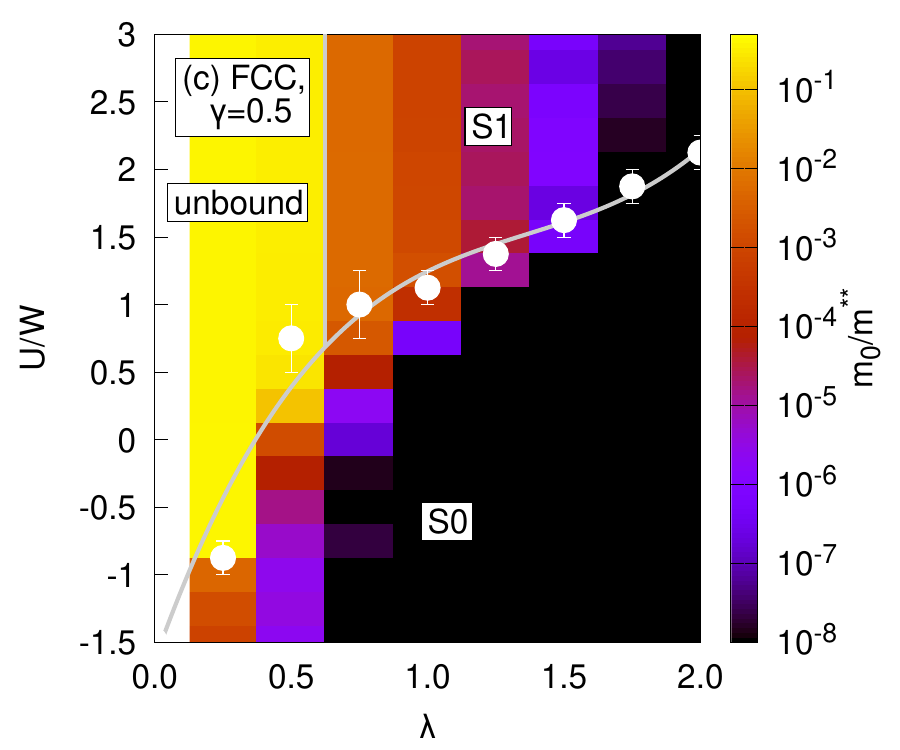}
\includegraphics[width=0.3\textwidth]{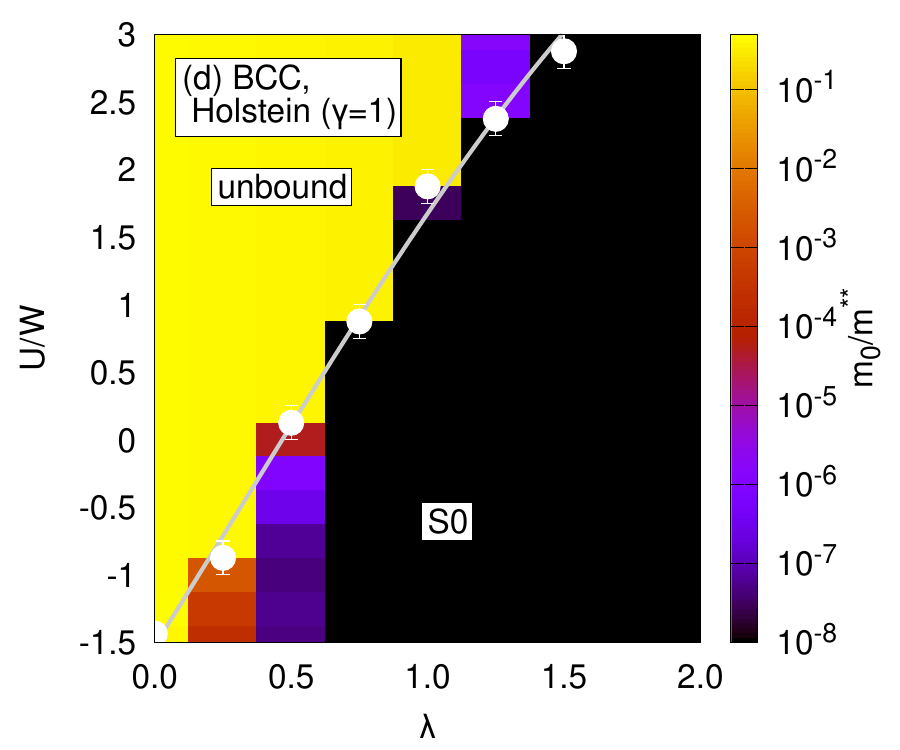}
\includegraphics[width=0.3\textwidth]{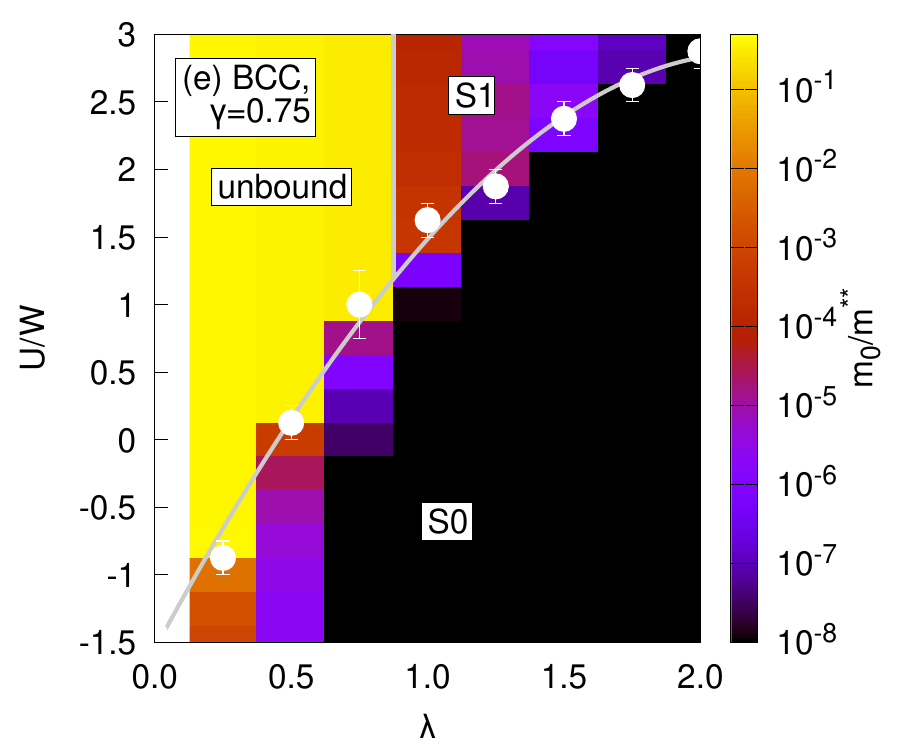}
\includegraphics[width=0.3\textwidth]{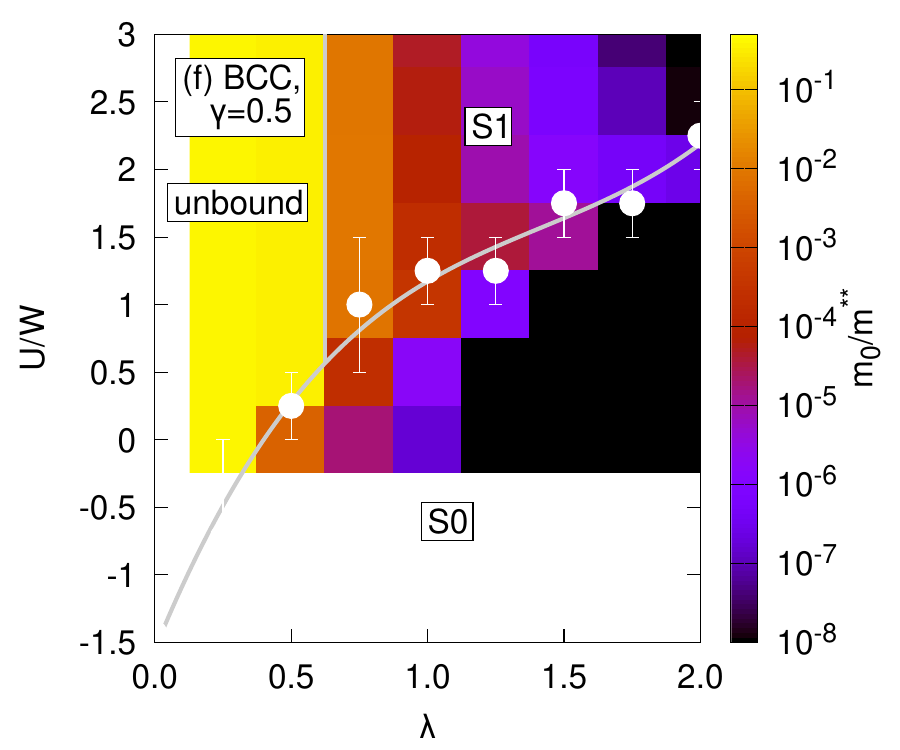}
    \caption{Inverse mass of the bipolaron in units of the electron mass, $m_{0}/m^{**}$. Calculations are in the adiabatic limit, $\hbar\omega = t$, which corresponds to $\hbar\omega = W/12$ for BCC and $W/16$ for FCC cases. A sharp change is found at the boundary of the S0 region of the binding diagram.}
    \label{fig:invmass}
\end{figure}

We show the inverse mass of the bipolarons in Fig. \ref{fig:invmass}. The S0 bipolaron is typically much heavier than the S1 bipolaron, and is typically only light close to the edge of the S0 region. For the $\hbar\omega=t$ cases, the S1 bipolaron is lighter for the $\gamma = 0.5$ case and heavier for the $\gamma = 0.75$ case. Generally the Holstein bipolaron is heaviest. We note that for the FCC cases, $\hbar\omega = t$ is further into the adiabatic regime, since $\hbar\omega/W$ is smaller due to the larger band width. 

\begin{figure}
\includegraphics[width=0.3\textwidth]{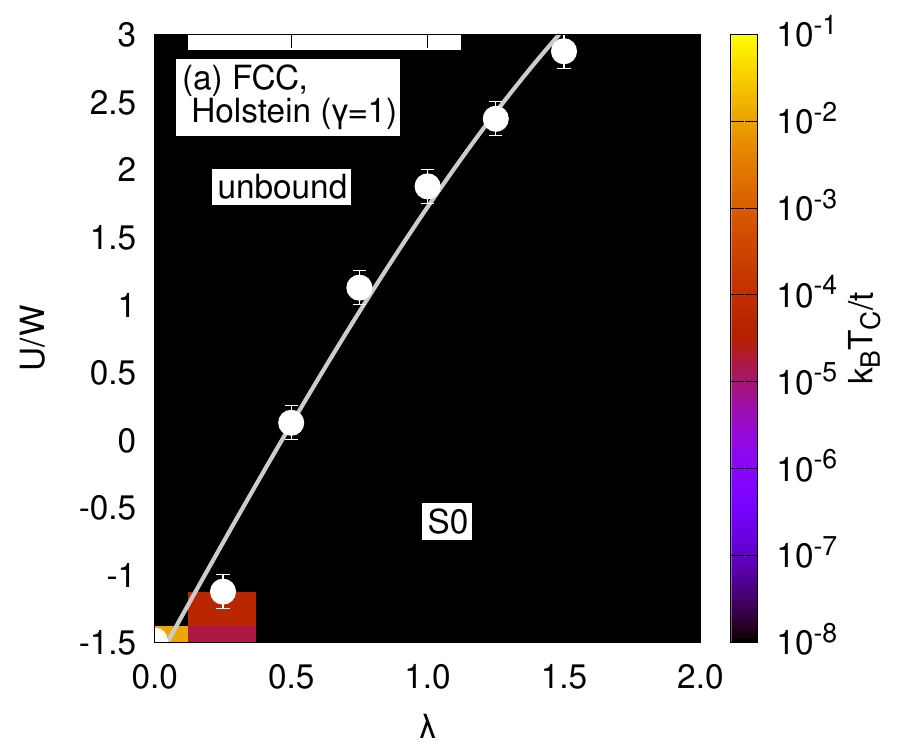}
\includegraphics[width=0.3\textwidth]{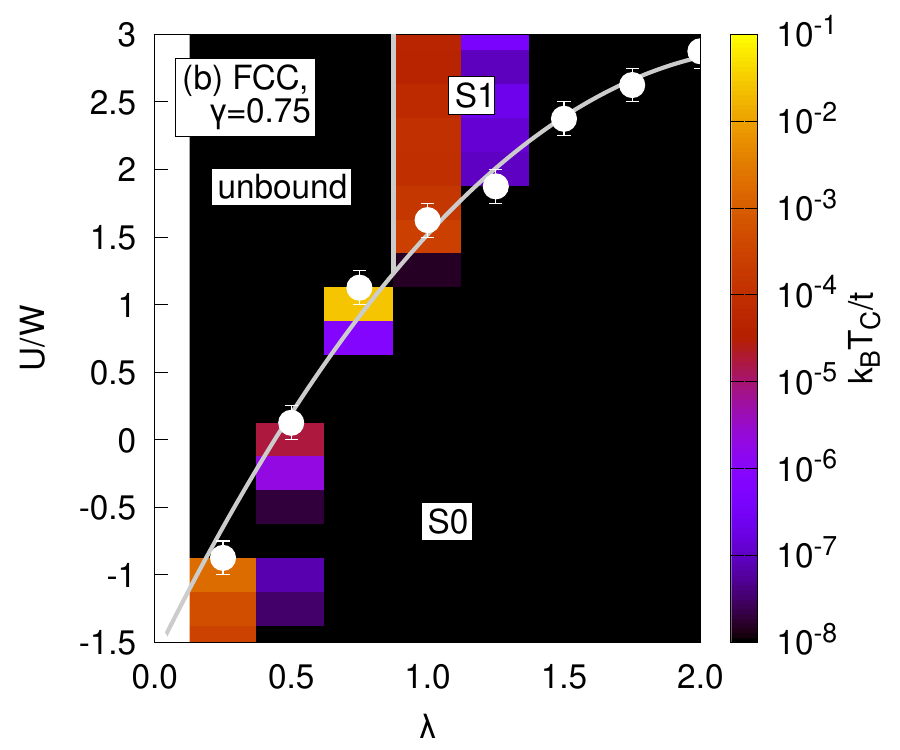}
\includegraphics[width=0.3\textwidth]{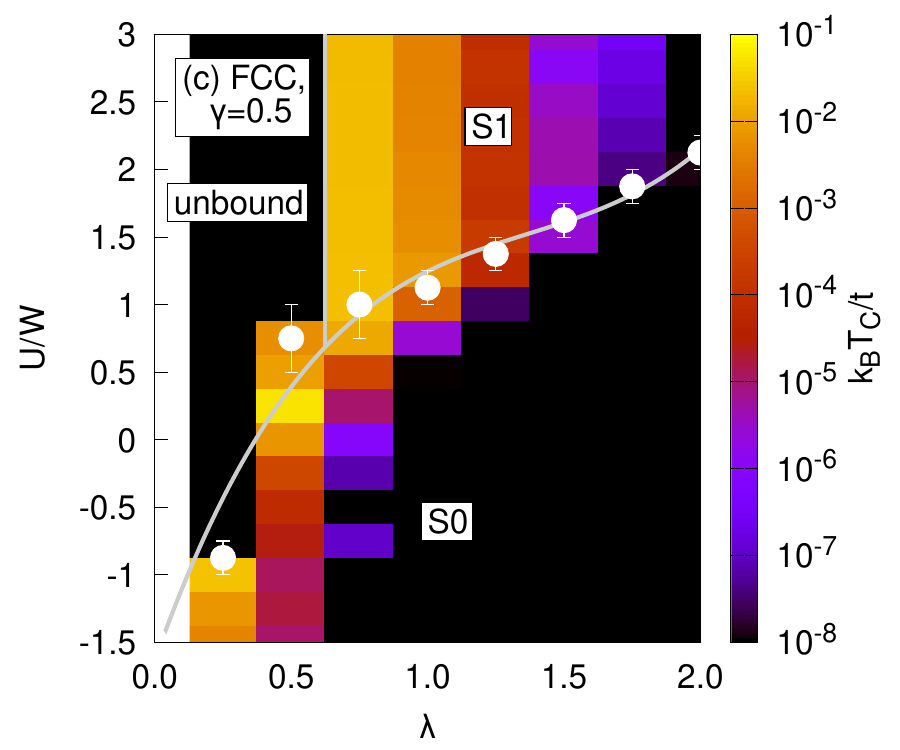}
\includegraphics[width=0.3\textwidth]{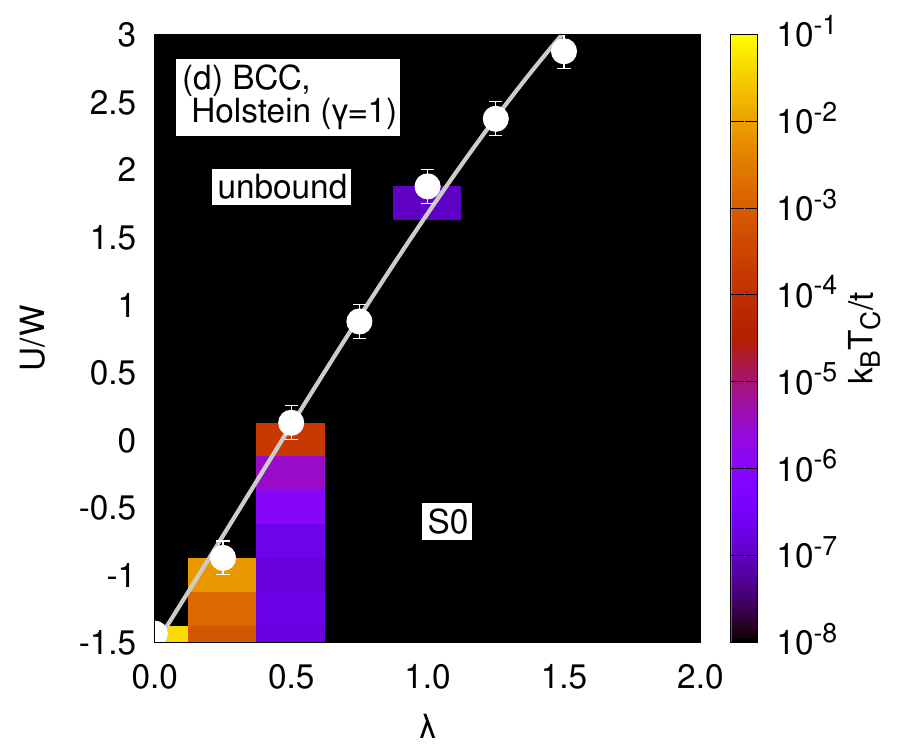}
\includegraphics[width=0.3\textwidth]{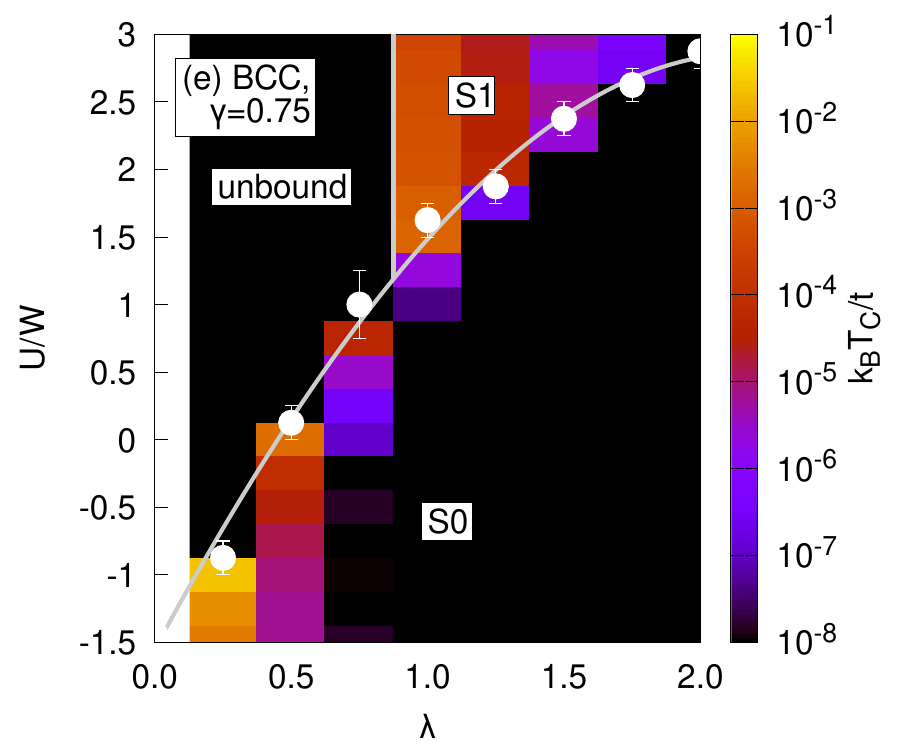}
\includegraphics[width=0.3\textwidth]{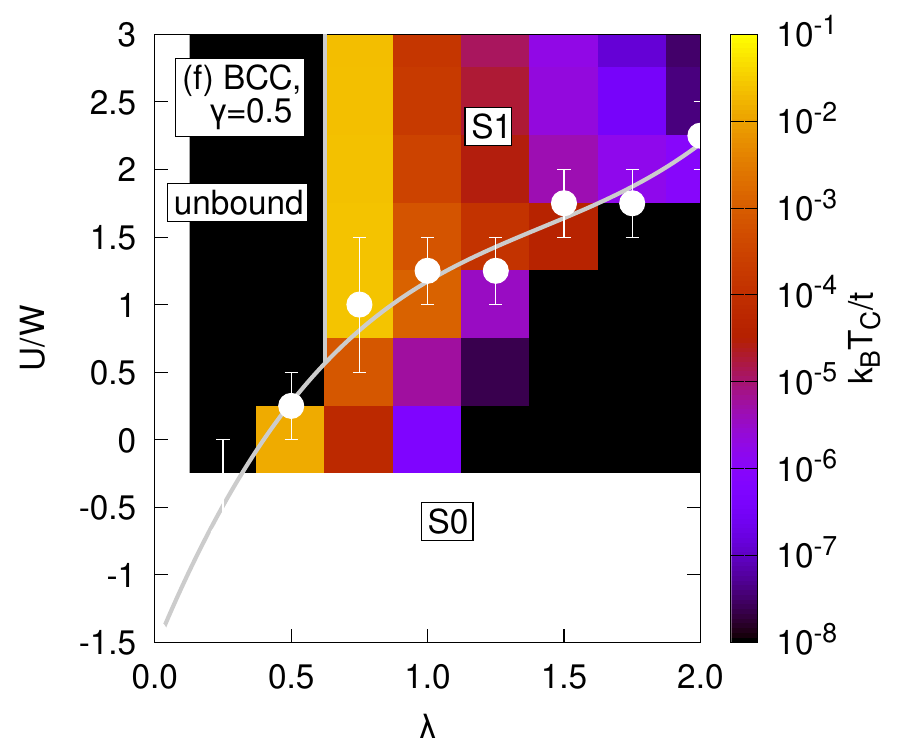}
    \caption{Close packing transition temperature ($T^{*}$) estimates in the adiabatic limit (from Eq. \ref{eqn:twoprime}) $\hbar\omega = t$, which corresponds to $\hbar\omega = W/12$ for BCC and $W/16$ for FCC cases. The FCC bipolaron has a higher maximum transition temperature than the BCC case for $\gamma=0.5$, in spite of the larger polaron mass. For $\gamma=0.75$, the BCC case has a higher transition temperature. There is a small region of higher $T^{*}$ associated with large bipolarons at low $U$ and $\lambda$ for $\gamma=1$ (Holstein model) that is outside the BEC regime.}
    \label{fig:tc}
\end{figure}

Figure \ref{fig:tc} shows transition temperature estimates. The FCC bipolaron has a higher maximum transition temperature than the BCC case for $\gamma=0.5$, in spite of the larger polaron mass. For $\gamma=0.75$, the BCC case has the higher transition temperature. There is a small region of large S0 bipolaron with higher estimated transition temperature in the Holstein case at small $U$ and $\lambda$. Since the pairs are large for the S0 bipolaron the system approaches the BCS rather than BEC limit in for $\lambda\rightarrow 0$. We note again that the ratio $\hbar\omega/W$ is smaller in the FCC than the BCC case so that there is more phonon creation and this may lead to a lowering of the relative transition temperature. In spite of this, the transition temperature is slightly higher for the FCC case with $\gamma=0.5$ over parts of the parameter space containing the S1 bipolaron.

\begin{figure}
\includegraphics[width=0.45\textwidth]{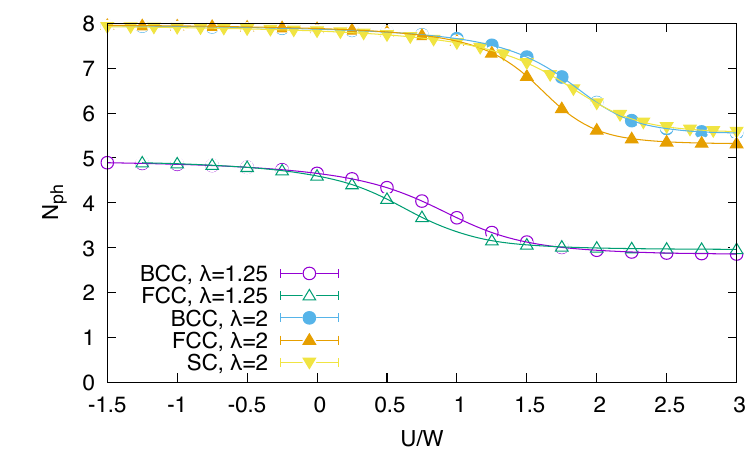}
\includegraphics[width=0.45\textwidth]{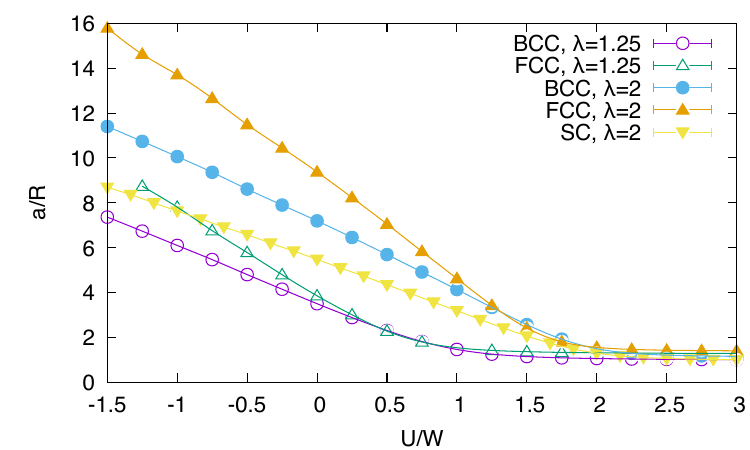}
\includegraphics[width=0.45\textwidth]{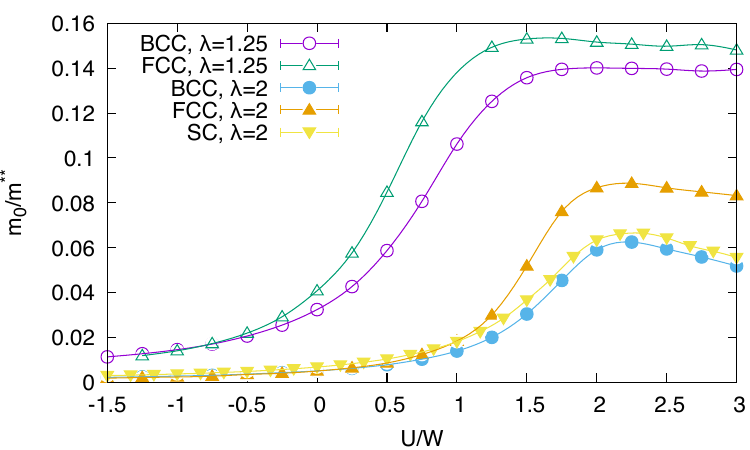}
\includegraphics[width=0.45\textwidth]{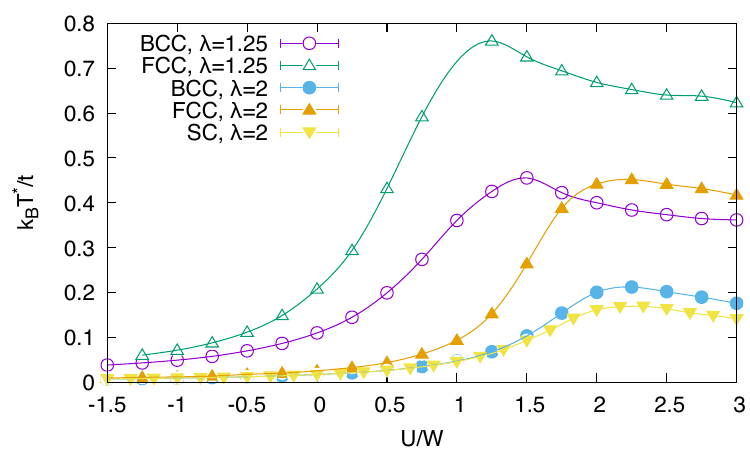}
    \caption{Inverse mass of the FCC and BCC bipolarons at large $\lambda$ and intermediate phonon frequency $\hbar\omega = W$. The FCC bipolaron is lighter (superlight) in this limit, with the S1 bipolaron much lighter than the S0 bipolaron in all cases. The small peak in inverse mass for $\lambda=2$ and BCC lattice (and to a lesser extent for the FCC lattice) corresponds to the light bipolaron case where intra- and inter-site effective interactions are of similar size on the boundary between S0 and S1 bipolarons. For larger phonon frequency, the number of phonons changes more slowly at the boundary between S0 and S1 bipolarons. The close packing transition temperature (from Eq. \ref{eqn:twoprime}) increases monotonically with $U$. As $\lambda$ increases, the percentage difference between the BCC and FCC cases becomes larger as the superlight regime is approached. For comparison, we also include data for the simple cubic (SC) lattice at $\lambda=2$ (see Ref. \cite{davenport2012}), for which all properties are very similar to the BCC case, $a/R$ differs primarily due to the difference in the near-neighbour distance, as does $T^{*}$. Overall, $T^{*}$ is much higher for the FCC case. For each data point, the error bars are displayed as 3 standard deviations. Where the error bars are not visible it means that they are too small to be apparent on the scale of the plot. Lines are a guide to the eye.}
    \label{fig:largeomegabonca}
\end{figure}

\subsection{Intermediate adiabaticity}

Figure \ref{fig:largeomegabonca} shows the inverse mass, number of phonons, inverse radius and estimated transition temperature of the FCC and BCC bipolarons at large $\lambda$ and intermediate adiabaticity, $\hbar\omega = W$, for $\gamma=0.5$. The FCC bipolaron is lighter than the BCC case by around 50\% in this limit, with the S1 bipolaron much lighter than the S0 bipolaron in all cases. At the larger $\lambda$, the bipolaron is slightly lighter at the boundary between the S0 and S1 bipolarons (between small and large $U$), as a manifestation of the crawler resonance of the bipolaron which can be identified in the anti-adiabatic limit (where the on-site and intersite bipolaron configurations have similar energy) as in Fig. 1(B). The superlight mechanisms that we have previously discussed have been established for the anti-adiabatic limit of high phonon frequency. We therefore expect systems with large phonon frequencies (typically the case in systems with low kinetic energies) to have more signatures of this superlight mechanism. At phonon frequency $\hbar\omega=W$, the number of phonons associated with the bipolaron shows a smoother crossover between the S0 and S1 states. The number of phonons is much smaller than for the $\hbar\omega = t$ case reflecting the additional energy carried with the phonons in the bipolaron cloud.  The inverse radius displays similar behaviour to the $\hbar\omega=t$ case. For comparison, we also include data for the simple cubic (SC) lattice at $\lambda=2$ (see Ref. \cite{davenport2012}), for which $N_{\rm ph}$ and $m_{0}/m^{**}$ are very similar. $a/R$ differs primarily due to the difference in the near-neighbour distance, as does $T^{*}$. Overall, $T^{*}$ is much higher for the FCC case. 

Finally, the close packing transition temperature is shown in Panel \ref{fig:largeomegabonca}(d). The estimated close packing transition temperature (from Eq.~(\ref{eqn:twoprime})) is much higher in the FCC case than the BCC case (and SC case), and the ratio of the transition temperatures for the two cases increases with the electron-phonon coupling strength as the bipolaron becomes more tightly bound between sites and the superlight effect becomes more important. The bipolarons become lighter at the point where the on-site and intersite effective interactions become similar, corresponding to a peak in the estimated transition temperature associated with $U'\sim V'$ in all cases.

\section{Discussion and conclusions}\label{sec:conclusions}

Using continuous time path-integral QMC simulations, we have studied the bipolaron in the Hubbard--Holstein and extended Hubbard--Holstein models, which contain local Coulomb repulsion in the presence of phonon-mediated attractive interactions. We have focused on pairing in BCC and FCC lattices. We used $N_{\rm ph}$ and $R$ to determine the binding (phase) diagram. Mass was calculated using twisted boundary conditions. Transition temperatures were estimated. A benefit of the QMC simulation is that we can examine the effect of phonon retardation on the properties of bipolarons. When $\hbar\omega\ll W$, the retardation effects are significant. 

We found that intersite interactions lead to qualitatively different pair properties: strongly bound EHHM bipolarons form both S0 and S1 pairs whereas HHM bipolarons are only found to have onsite characteristics. Specifically, at large $U$ and $\lambda\gtrsim 1$, HHM and EHHM bipolarons have qualitatively different behaviours: intersite EHHM bipolarons can be well bound and stable at large $U$ above a critical electron-phonon coupling, $\lambda\ge \lambda_{C}$, due to the near-neighbour phonon attraction. The value of $\lambda_{C}$ decreases with decreased $\gamma$.

Overall, we found no qualitative difference in bipolaron binding diagrams for BCC and FCC lattices for $\hbar\omega=t$. S1 bipolarons at intermediate electron-phonon coupling tend to be lighter on the FCC lattice for larger intersite coupling and heavier for lower intersite coupling. In the Holstein case, the FCC bipolaron is much heavier and as $\gamma$ decreases, the FCC is lighter. We note that there are also fulleride systems with simple cubic (SC) lattices. We have previously studied the SC version of the extended Holstein Hamiltonian \cite{davenport2012} for $\hbar\omega = W$.

On increase of phonon frequency, retardation effects become less important and the effective interaction Hamiltonian becomes more Hubbard like. We observed that for $\hbar\omega=W$, signatures of superlight states (where the pair moves by first-order hops) start to emerge, and lower masses are found for FCC lattices, especially at large $\lambda$. We also see evidence of superlight behaviour for pairs on BCC lattices of the type shown in Fig. \ref{fig:schematic}(B) for the specific case where $U' \sim V'$. Estimated transition temperatures in this case are larger for bipolarons on FCC lattices.

As further work, study of the EPIs in A$_3$C$_{60}$ compounds would be interesting to establish whether any longer range EPIs that could lead to superlight effects are present in fullerides. We briefly discuss our results in the context of parameters relating to the A$_3$C$_{60}$ compounds. These materials display superconductivity for both BCC and FCC lattice structures. Their bandwidth $W\sim 0.5 \mathrm{eV}$. The Coulomb repulsion in A$_3$C$_{60}$ superconductors is also large with $1.5\lesssim U/W \lesssim 2.5$ and typical values for the nearest-neighbour Coulomb potentials range from about $0.5- 0.8W$ \cite{gunnarsson1996mott,gunnarsson2004alkali,lof1992band,martin_and_ritchie_1993,antropov1992coulomb,lof1992correlation,bruhwiler1993auger}. The high-frequency intramolecular vibron modes, $\hbar\omega/W \sim 0.1-0.4$, play a key role in the high critical temperatures. Electron-phonon coupling involves multiple bands (of a Jahn--Teller type), but single band models may be used to explore some of the physics involved with the electron-phonon interactions. The strongest electron-phonon couplings are intra-molecular and involve coupling between electrons and Einstein modes. They are estimated from experiments to be of order $\lambda\sim 0.5-1.0$. The different types of electron-phonon interaction present in the fullerides are described in Ref. \cite{gunnarsson2004alkali}. Intersite electron-phonon interactions mediated by neighbouring $C_{60}$ sites or alkali metal sites would be required to obtain S1 bipolarons. Inter-molecular interactions between electrons on C$_{60}$ molecules are weak. The strength of interactions between electrons on C$_{60}$ molecules and alkali metal ions is unclear (with DFT calculations suggesting $\lambda\lesssim 0.1$, but other calculations estimating higher interactions). Determining the strength of these interaction in more detail would be useful to establish if the they are sufficiently strong to bind lighter S1 bipolarons.

Additional QMC calculations including next-nearest neighbour hopping would also be of interest for study of the BCC lattice, since the intersite spacings of near-neighbour and next-nearest neighbour are similar, potentially leading to similar hoppings. The presence of similar hopping strengths can lead to superlight effects in the anti-adiabatic limit, and the results in this paper indicate that superlight effects can persist into the intermediate adiabicity regimes  \cite{adebanjo2024}. In practice, these next-nearest neighbor interactions are likely to decrease mass and increase transition temperatures due to the 1st order processes for bipolaron motion enabled by those extra hops. We discussed the process in the anti-adiabatic limit in our recent paper \cite{adebanjo2024}. QMC simulation of the effects of the next-nearest neighbour hoppings is beyond the scope of the current paper as it would require significant algorithmic development with additional 3 kink updates involving both nearest and next-nearest neighbour hopping and is reserved for a future publication.

\bibliographystyle{unsrt}
\bibliography{References}

\end{document}